\documentclass[journal]{IEEEtran}
\usepackage{cite}

\ifCLASSINFOpdf
\else
\fi
\usepackage{amsmath}
\usepackage{array}

\ifCLASSOPTIONcompsoc
  \usepackage[caption=false,font=normalsize,labelfont=sf,textfont=sf]{subfig}
\else
  \usepackage[caption=false,font=footnotesize]{subfig}
\fi
\usepackage{xcolor}
\usepackage[flushleft]{threeparttable}
\usepackage{siunitx}
\usepackage{graphicx}
\usepackage{amsmath}
\usepackage{mathtools}
\usepackage{comment}
\usepackage{multirow} 
\usepackage{color}
\usepackage[normalem]{ulem}

\begin{document}

\title{Perceptual Video Quality Prediction Emphasizing Chroma Distortions}
%
%
%

\author{Li-Heng~Chen,
        Christos~G.~Bampis,
        Zhi~Li,
        Joel~Sole,
        and~Alan~C.~Bovik,~\IEEEmembership{Fellow,~IEEE}
\thanks{L.-H.~Chen and A.~C.~Bovik are with the Department of Electrical and Computer Engineering, University of Texas at Austin, Austin, TX, 78712 USA (email:lhchen@utexas.edu, bovik@ece.utexas.edu).}
\thanks{C.~G.~Bampis, Z.~Li, and J.~Sole are with Netflix Inc. Los Gatos, CA, 95032 USA (email:christosb@netflix.com, zli@netflix.com, jsole@netflix.com).}
\thanks{This work is supported by Netflix. Part of this work was carried out during L.-H. Chen's summer internship at Netflix.}
}

%
%

\markboth{preprint, under review}%
{Shell \MakeLowercase{\textit{et al.}}: Bare Demo of IEEEtran.cls for IEEE Journals}
%



\maketitle

\begin{abstract}
Measuring the quality of digital videos viewed by human observers has become a common practice in numerous multimedia applications, such as adaptive video streaming, quality monitoring, and other digital TV applications. Here we explore a significant, yet relatively unexplored problem: measuring perceptual quality on videos arising from both luma and chroma distortions from compression. Toward investigating this problem, it is important to understand the kinds of chroma distortions that arise, how they relate to luma compression distortions, and how they can affect perceived quality. We designed and carried out a subjective experiment to measure subjective video quality on both luma and chroma distortions, introduced both in isolation as well as together. Specifically, the new subjective dataset comprises a total of 210 videos afflicted by distortions caused by varying levels of luma quantization commingled with different amounts of chroma quantization. The subjective scores were evaluated by 34 subjects in a controlled environmental setting. Using the newly collected subjective data, we were able to demonstrate important shortcomings of existing video quality models, especially in regards to chroma distortions. Further, we designed an objective video quality model which builds on existing video quality algorithms, by considering the fidelity of chroma channels in a principled way. We also found that this quality analysis implies that there is room for reducing bitrate consumption in modern video codecs by creatively increasing the compression factor on chroma channels. We believe that this work will both encourage further research in this direction, as well as advance progress on the ultimate goal of jointly optimizing luma and chroma compression in modern video encoders.
\end{abstract}


\IEEEpeerreviewmaketitle

\section{Introduction}
\IEEEPARstart{M}{odeling} the human perception of video quality has become a crucial research problem owing to the tremendous boom of shared and streaming video content, accessible mobile video devices, and diverse video services such as Netflix, Facebook, Youtube, Hulu, and so on. Perceptual optimization of multimedia workflows is important, since humans are the direct consumers of visual information. Videos have become the dominant portion of Internet traffic, and it is predicted that pictures and videos will comprise 80\% of the moving bits in broadband and mobile networks in the near future \cite{ciscovni}. Given this tremendous, growing, network bandwidth demand, improving the rate-distortion performance of video compression in perceptual ways has become a critical issue, with the potential for significant social, ecological, and economic impact.

Because of significant efforts applied to developing algorithms that can accurately predict perceptual video quality, many promising models have been demonstrated. One successful example is the trained Netflix video quality prediction model called Video Multimethod Assessment Fusion (VMAF) \cite{ZliVMAF16, ZliVMAF18}, which is used to optimize a significant percentage of Internet video traffic at a level only matched by SSIM \cite{WangBSS04}. However, most practically deployed video quality models, including VMAF, only extract visual information from the video luma channel, disregarding color entirely. Yet, neglecting color/chroma components may be detrimental to achieving the most accurate evaluations of video quality. As a practical example, one may intentionally manipulate the chroma fidelity of a video: before encoding a pristine source video, it is common to decimate the chroma components to reduce bitrate consumption, with little impact on predicted or perceived quality. Of course, chroma quality features have been used in some simple ways. For example, averaging the PSNR values across luma and chroma channels has been used in multimedia tools such as FFmpeg. Similarly, the Moving Picture Experts Group (MPEG) community often evaluates the performance of video codecs using a weighted mean of Bjøntegaard-Delta bitrates (BD-rate) \cite{BDRate01} across per-channel PSNR values. However, rather than taking human perception into account, the linear weights were empirically derived as a proportion of color channels defined by pixel format. 

\begin{figure}[!t]
	\centering
	\footnotesize
	\renewcommand{\tabcolsep}{1.5pt} 
	\def\imgwid{0.234\textwidth}
	
	\begin{tabular}{cc}
    \multicolumn{2}{c}{YCbCr displayed with UTU-R BT.709 color conversion} \\
    \includegraphics[width=\imgwid]{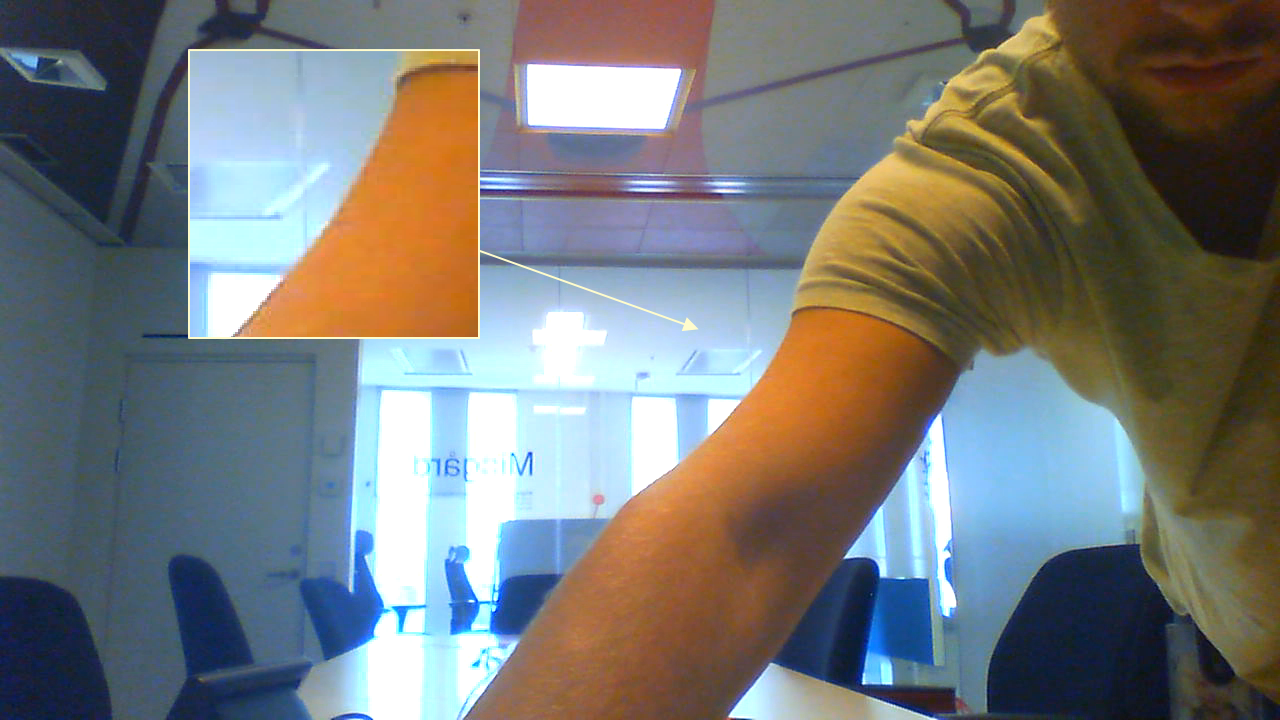} &
    \includegraphics[width=\imgwid]{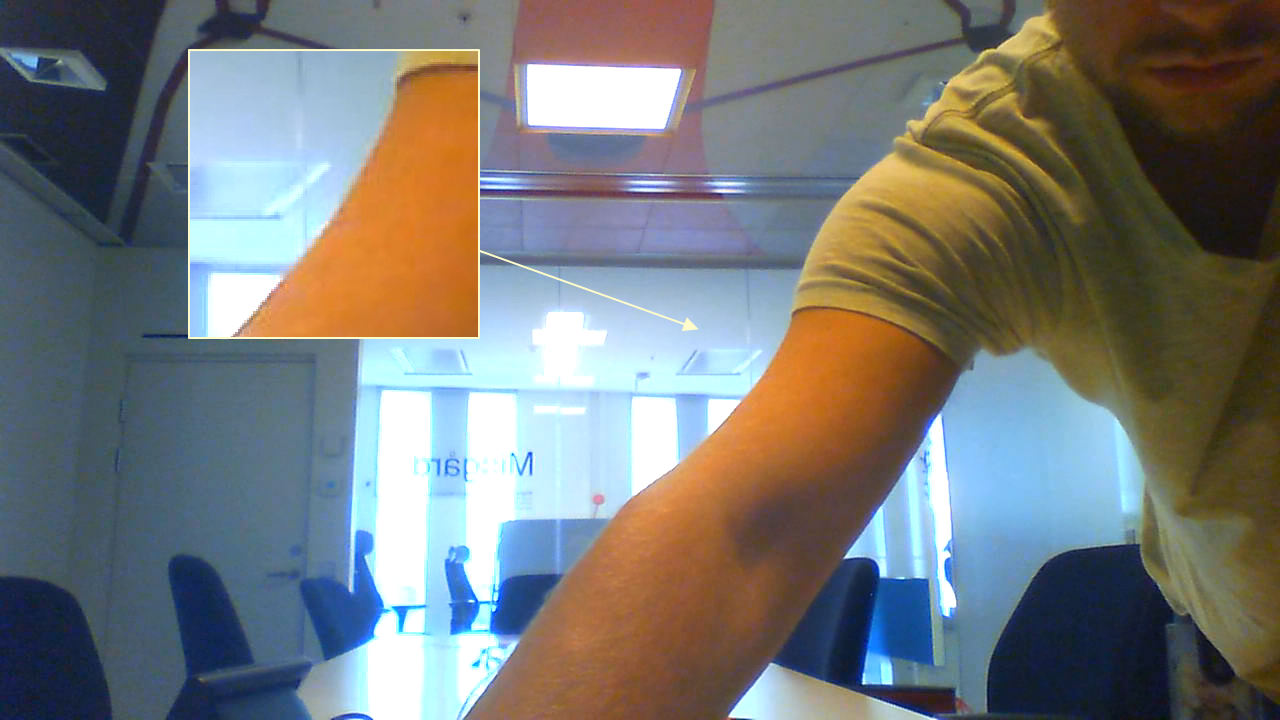} \\
    \multicolumn{2}{c}{Cb channel displayed in gray scale} \\
    \includegraphics[width=\imgwid]{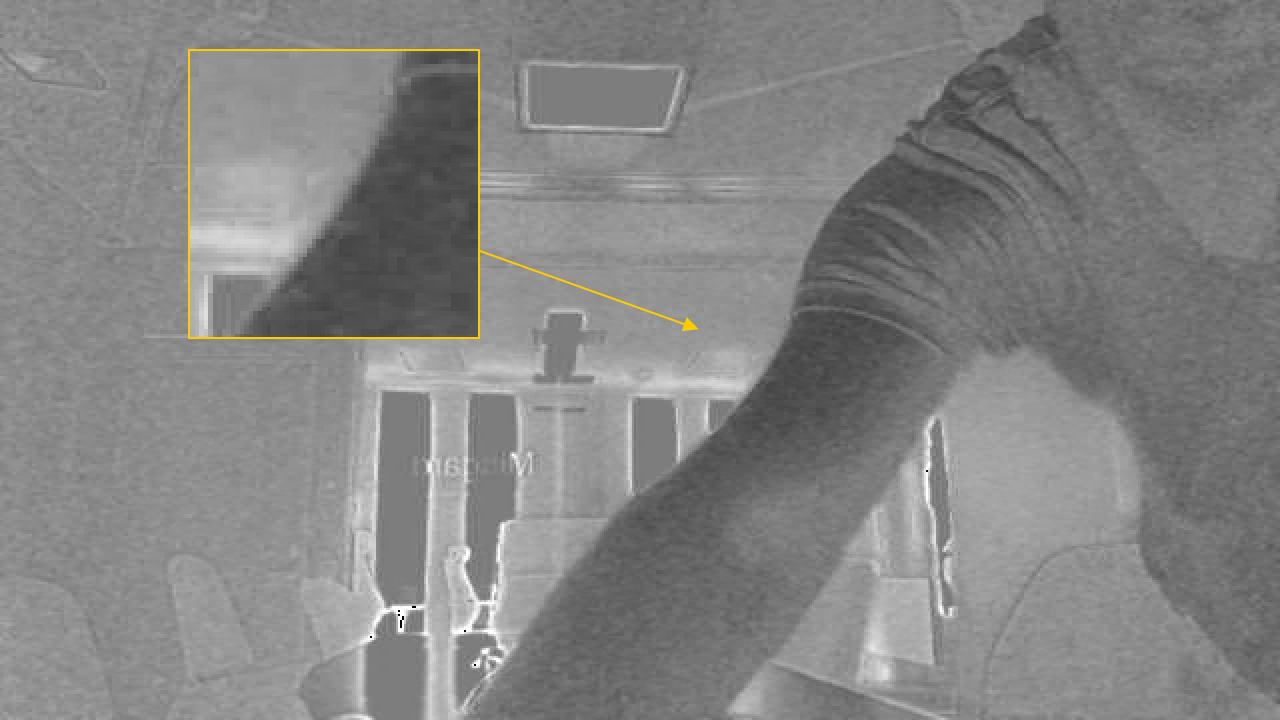} &
    \includegraphics[width=\imgwid]{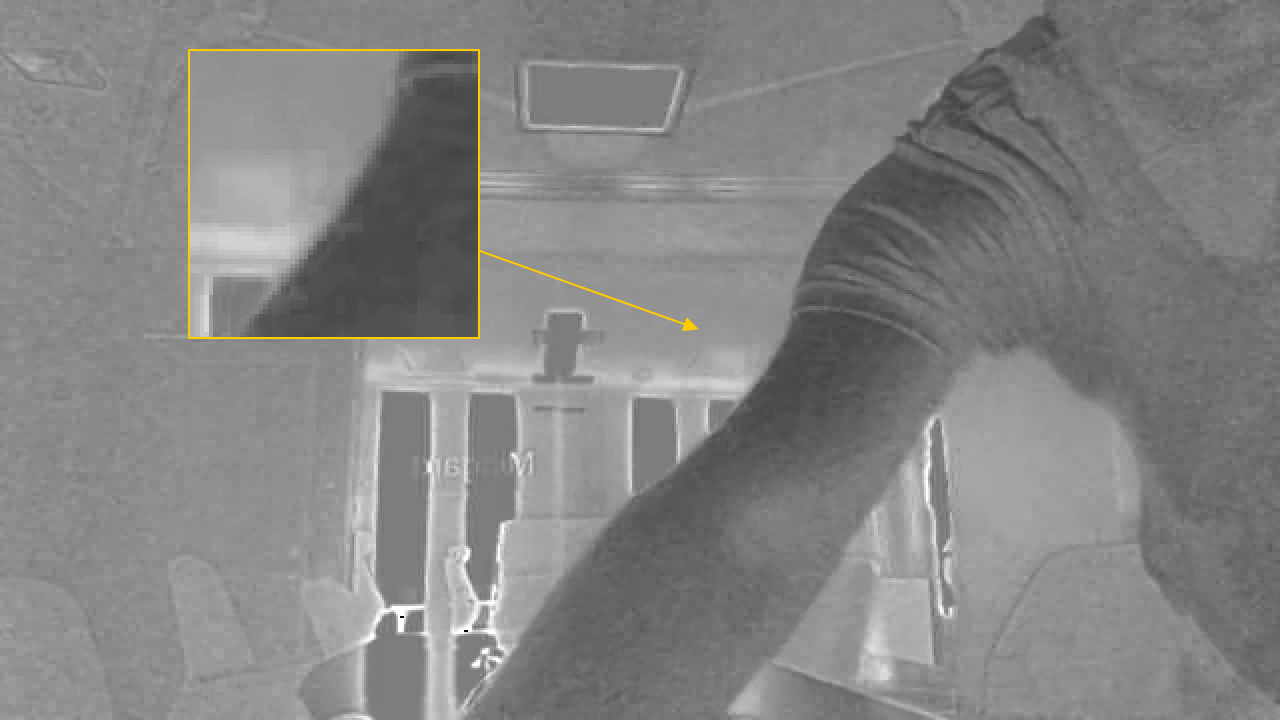} \\
    \multicolumn{2}{c}{Cr channel displayed in gray scale} \\
    \includegraphics[width=\imgwid]{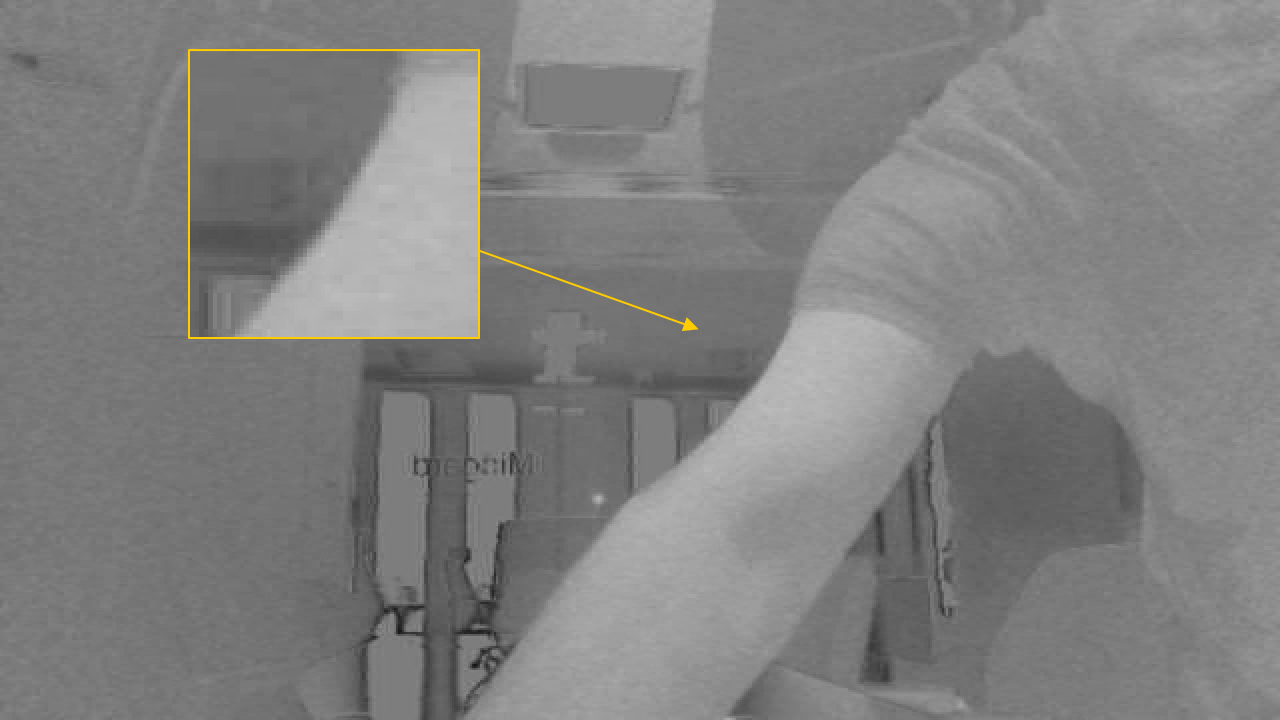} &
    \includegraphics[width=\imgwid]{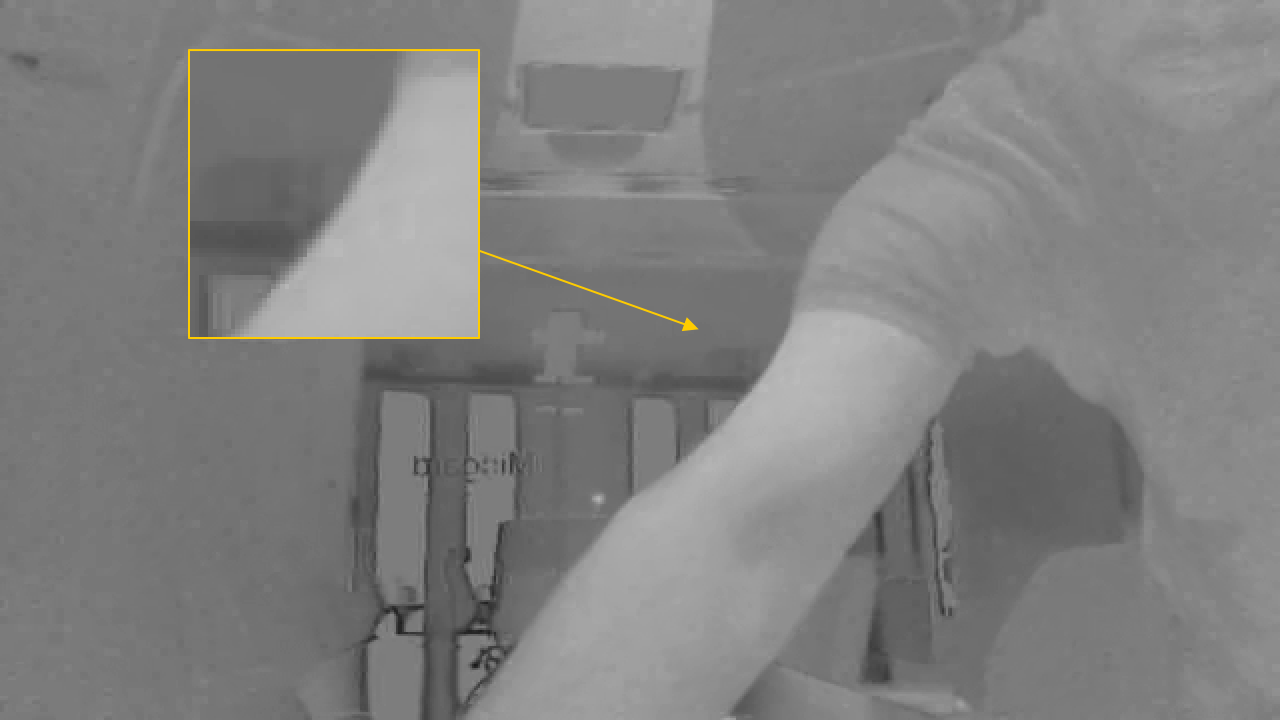} \\
    (a) \textit{chroma\_qp\_offset}~=~0 & (b) \textit{chroma\_qp\_offset}~=~$K$
	\end{tabular}
	\caption{A bitrate-saving example of encoding test sequence \textit{dark720p\_120f} using HEVC. The two videos were encoded at a fixed QP of 17 with different \textit{chroma\_qp\_offset} settings, resulting in (a) $\text{PSNR}_\text{Cb}=47.77$ db; $\text{PSNR}_\text{Cr}=49.01$ db; $\text{bitrate}=20631.65$ Kbps, and (b) $\text{PSNR}_\text{Cb}=39.50$ db; $\text{PSNR}_\text{Cr}=41.92$ db; $\text{bitrate}=13112.41$ Kbps }
	\label{fig:cqp_demo}
\end{figure}

One potentially high-impact application of perceptually quantifying chroma distortion is the optimization of video compression. For example, in Fig. \ref{fig:cqp_demo} we applied several different quantization factors settings on noisy chroma channels. Here, distortions of Cb and Cr caused by maximizing the \textit{chroma\_qp\_offset} parameter to $K$, thereby increasing quantization of Cb and Cr, are clearly visible. However, while the visual differences caused by the two settings are subjectively and objectively noticeable when displaying chroma channels independently, it is important to observe that these considerable defects in chroma do not alter the visual quality when YCbCr are displayed together. Yet, in terms of bit consumption, the encoded video in Fig.~\ref{fig:cqp_demo}(b) only occupies $6.5$ Mbytes, which is $36\%$ less than the video in Fig.~\ref{fig:cqp_demo}(a). Unfortunately, the potentially significant benefits of trading off chroma fidelity against bitrate cannot be captured by the conventional per-channel PSNR, and cannot contribute to improved PSNR-based rate-distortion optimization schemes.

\begin{figure}[!t]
  \centerline{
  \includegraphics[width=1.04\columnwidth]{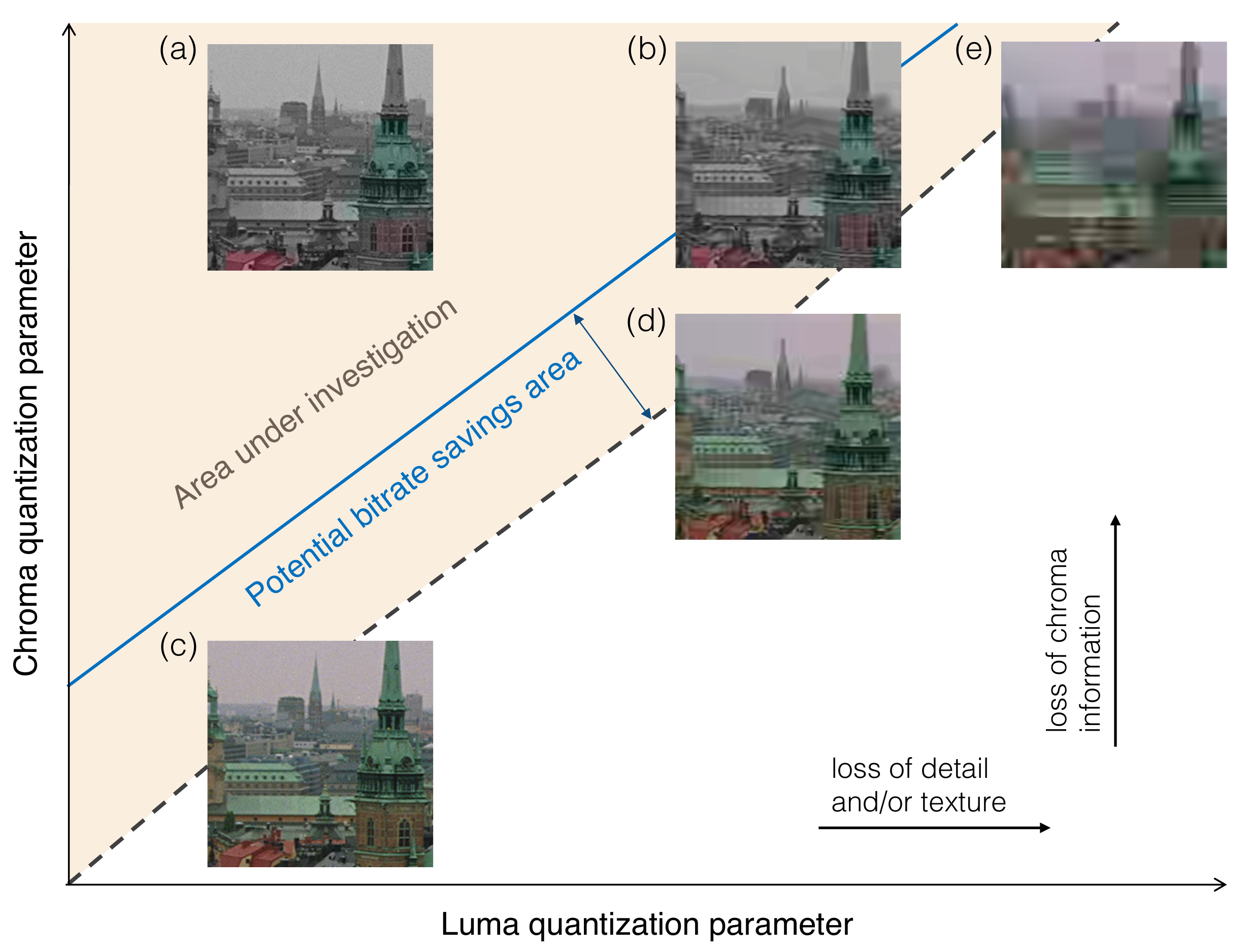}
  }
  \caption{Conceptual overview of this paper. Each patch shows distortion from a specific combination of luma and chroma quantization. The black dotted line denotes the default setting \textit{chroma\_qp\_offset}$=0$. The area between the black dotted diagonal and the blue solid diagonal illustrates where modern codecs are currently allowed to vary quantization in chroma ($0\le\textit{chroma\_qp\_offset}\le K$).}
  \label{fig:overview}
  \end{figure}

We summarize the main idea behind this work in Fig. \ref{fig:overview}. Most video encoders operate near the region defined by the black dotted diagonal line, where the amount of compression between chroma and luma is tightly coupled. Using a similar working assumption, most previous subjective quality databases and most quality models do not consider the perceptual effects of decoupling luma and chroma compression. In these scenarios, as shown in Figs. \ref{fig:overview}(c), (d) and (e), chroma artifacts are typically observed only when the luma quantization factor is high enough (as in Fig. \ref{fig:overview}(e)). Notably, the upper left region provides room for further bitrate savings, by exploiting heavier chroma quantization while fixing the luma quantization level, without coupling the two. Figure \ref{fig:overview}(a) shows an example of severely degrading chroma, while luma quantization is kept to a much lower level. As a result, it exhibits significant loss of chroma information (desaturation), but the scene structure remains intact. Moving across Fig. \ref{fig:overview} in the horizontal direction, luma quantization increases, until significant loss of detail and texture are observed (as in Fig. \ref{fig:overview}(b)). Notably, the bottom-right of Fig. \ref{fig:overview} is a region where luma is more severely quantized than chroma, an approach that should be less efficient from the rate-distortion point of view, given that human perception is more sensitive to loss of detail.

Motivated by the significant potential of perceptual chroma quality prediction and optimization, we have attempted to improve the current VMAF model by accounting for the perceptual quality attributes of the chroma components of videos. To help us advance progress in this direction, we built a new chroma distortion-specific video quality database, on which we conducted a subjective quality study to better understand how different levels of compression distortion in chroma affect quality perception, and how they relate to luma distortions. This study seeks to examine ways of exploiting the upper-left area of Fig. \ref{fig:overview}, to better understand the perceptual tradeoffs that should mediate luma and chroma compression. Using the collected data, we developed and tested new chroma quality-aware features which can be used to improve existing learning based video quality predictors. Moreover, we explore the potential of using color-sensitized video quality predictors, like the improved VMAF model introduced here, to improve the perceptual optimization of compression.

Moving forward, section II reviews related literature and motivates the need for a new dataset. Section III details the new database's source contents, the creation of chroma distortions, and the perceptual study design and outcomes. Section IV discusses data analysis of the subjective study results. Section V explores different chromatic features and appropriate design methodologies for constructing quality prediction models, while section VI offers an analysis of the performance of a variety of objective VQA algorithms on the new database. Finally, section VII concludes with a discussion of future directions of research.

\section{Background}
We begin with a background review of studies related to subjective video quality. Following that, currently available objective video quality assessment algorithms also discussed.

\subsection{Subjective Video/Picture Quality Databases}
Numerous Video Quality Assessment (VQA) datasets have been designed over the past decade. The early LIVE VQA Database \cite{Seshadrinathan2010} and the LIVE Mobile Video Quality Database \cite{Moorthy2012} remain among the most widely used public-domain video quality databases. These databases model post-acquisition distortions, such as H.264 compression and transmission errors, to match real world scenarios. Other similar databases, of videos having larger resolutions, deeper bit-depths, or encoded by more advanced codecs have been proposed, including the CSIQ-VQA Database \cite{Vu2014}, the SHVC Database \cite{SHVC_DB}, and the VQEG Database \cite{VQEG_DB}. We refer the reader to \cite{winkler2012analysis,WinklerWebsite} for a comprehensive analysis and collection of publicly available databases through 2012. In recent years, more video quality databases \cite{Cheon2018,ZhangBVIHD2018,Ghadiyaram2019,RamachandraRao2019, yu2020ugcvqa, DYLee2020} have been proposed to address different scenarios. Although these databases have undoubtedly accelerated the development of VQA algorithms, generally distortions of the chroma channels were not specifically designed. In most cases, chroma distortion are hightly coupled with luma distortion.

To the best of our knowledge, we are aware of only a few studies of chromatic distortions. Shang \textit{et al.} \cite{Shang2019} introduced different levels of chromatic distortions than conventional codecs. Source videos were divided into three planes, then independently encoded at different quantization levels. The independently encoded channels were then combined into a single distorted video. Although this methodology produces combinations of Y/Cb/Cr components having different levels of distortions, it is not naturally produced by a video codec. Here, we instead propose a different framework whereby distorted videos are generated directly from an encoder, as we will discuss.

There are also two databases that studied \textit{image} quality, including certain distortions of chroma components: TID2013 \cite{Ponomarenko2015} comprises various color distortions in addition to 17 other distortion types. However, the four synthesized color distortions are peculiar and not likely to be present in practical situations. In \cite{Sinno2020}, Sinno \textit{et al.} studied the quality of billboard and thumbnail images displayed by streaming services. This database contains JPEG-distorted still pictures in 444/420 formats, representative of modern internet applications. They found that chromatic distortions, such as color bleeding or jaggies, were induced by chroma subsampling. However, these artifacts are often negligible, and are generally only noticeable when viewed in side-by-side pairwise comparisons. They tend to be even more subtle when appearing on video, where temporal masking effects often suppress faint distortions.

\subsection{Objective Video Quality Assessment Algorithms}
Another closely related topic, objective video quality assessment, has been a long-standing and fundamental research problem. Generally, video quality prediction models are classified as full-reference (FR), reduced-reference (RR), or no-reference (NR), based on the availability of the high-quality reference data. Here we only focus on the FR scenario, since it may be assumed that ground-truth data is available in our target applications. Beyond the popular SSIM \cite{WangBSS04,WangVQA2004} and MS-SSIM \cite{WangMSSSIM03} models, which are computationally simple, numerous other powerful perceptual models have also been proposed. These include the VIF index \cite{SheikhB06} which supplies most of the features in VMAF, the MOVIE index \cite{MOVIE2010}, ST-MAD \cite{Vu2011}, the VQM-VFD models \cite{Pinson2004, wolf2011, Pinson2014}, and many others \cite{Hekstra2002, Tao2007, Soundararajan2013, BampisSpeed2017, ManasaVQA2016, HuVQA2017, Yu2019}. There are a few still picture (IQA) models \cite{Lissner2013, Preiss2014} designed to tackle chroma distortions, which produce improvements on the color-distorted images in the TID2013 database.

Despite the commercial success of many perceptual quality predictors, the pixel-wise PSNR is still a mainstream model, especially in the testing of video codecs. Indeed, variants of PSNR have been used in the context of codec comparison. During the coding process three PSNR values PSNR$_\text{Y}$, PSNR$_\text{Cb}$ and PSNR$_\text{Cr}$ are obtained. They are usually combined to produce PSNR$_\text{611}$ \cite{JCTVCH001212} or PSNR$_\text{411}$\footnote[1]{used in libavfilter of ffmpeg} as the final quality score:
\begin{equation}
\text{PSNR}_\text{k11} = \frac{\text{k}\cdot\text{PSNR}_\text{Y} + \text{PSNR}_\text{Cb} + \text{PSNR}_\text{Cr}}{\text{k}+2}.
\end{equation}
PSNR-HVS-M \cite{psnrhvsm07}, perceptually-oriented extension of PSNR, has been integrated into the Daala codec \cite{Valin2016} for performance evaluation. A recently proposed color-sensitivity-based combined PSNR (CS-PSNR) \cite{Shang2019}, uses perceptually optimized weights on Y/Cb/Cr based on a subjective analysis of color sensitivity. The MSE weight for each channel is inversely proportional to just-noticeable unit area of a checkerboard pattern. However, this model might be limited by its straightforward linear fusion of per-channel MSE's. Again, chromatic information is often neglected, or naively exploited, both in the design of databases and in quality prediction algorithms.

\begin{table}[tp]
  \caption{Specification of QP for chroma (QP$_\text{c}$) as a function of the transitional value QP$_\text{i}$ in HEVC.}
  \centering
  \scalebox{1.0}{
    \begin{threeparttable}
  \begin{tabular}{|c|c|c|c|c|c|c|c|c|c|}
  \hline
  QP$_\text{i}$ & $<30$             &30&31&32&33&34&35&36 \\ \hline
  QP$_\text{c}$ & QP$_\text{i}$   &29&30&31&32&33&33&34 \\ \hline\hline
  QP$_\text{i}$ & $>43$             &43&42&41&40&39&38&37 \\ \hline
  QP$_\text{c}$ & QP$_\text{i}$-6 &37&37&36&36&35&35&34 \\ \hline
  \end{tabular}
\end{threeparttable} 
  }
  \label{chroma_qp_mapping}
\end{table}

\section{Subjective Experiment Design}
Next we describe the key characteristics of the subjective study that we conducted to capture human judgments of compression distortion on both luma and chroma channels.

\subsection{Generating Chroma Compression Distortions}
In most modern video coding standards, the quantization parameters (QP) for chroma components are not explicitly designated. Instead, it is derived from the QP value of the luma channel with a parameter \textit{chroma\_qp\_offset} (the syntax name may vary across different codec standards) that gives a certain degree of flexibility. For example, syntaxes \textbf{cb\_qp\_offset} and \textbf{cr\_qp\_offset} are defined in the High Efficiency Video Coding (HEVC) standard. The offsets are first clipped to the range $\left[-12,12\right]$, then added to the luma QP (denoted by QP$_\text{Y}$):

\begin{equation}\label{eq:transQP}
  \begin{split}
  \mathrm{QP_{i,Cb}} & = \mathrm{QP_Y}+\mathrm{clip_{[-12,12]}}(\mathbf{cb\_qp\_offset}), \\
  \mathrm{QP_{i,Cr}} & = \mathrm{QP_Y}+\mathrm{clip_{[-12,12]}}(\mathbf{cr\_qp\_offset}).
  \end{split}
\end{equation}
Then, QP$_\text{i,Cb}$ and QP$_\text{i,Cr}$ are mapped to the QP values for Cb and Cr:
\begin{equation}\label{eq:finalCQP}
  \begin{split}
  \mathrm{QP_{Cb}} & = f\left( \mathrm{QP_{i,Cb}} \right), \\
  \mathrm{QP_{Cr}} & = f\left( \mathrm{QP_{i,Cr}} \right), \\
  \end{split}
\end{equation}
where $f(.)$ is a nonlinear mapping function normally implemented as a look-up table. Table \ref{chroma_qp_mapping} describes this function in the HEVC standard. Consequently, a tremendous challenge in generating realistic chroma compression distortions is that the \textit{chroma\_qp\_offset} parameter in all of the video compression standards are limited to the range $[-12,12]$, hence one cannot directly produce compressed videos with arbitrary chroma quantization levels distinct from the luma quantization parameter. 

\begin{figure}[!t]
	\centering
	\footnotesize
	\renewcommand{\tabcolsep}{1.1pt} 
	\renewcommand{\arraystretch}{1.2} 
	\def\imgwid{0.235\textwidth}
	\begin{tabular}{cc}
    \includegraphics[width=\imgwid]{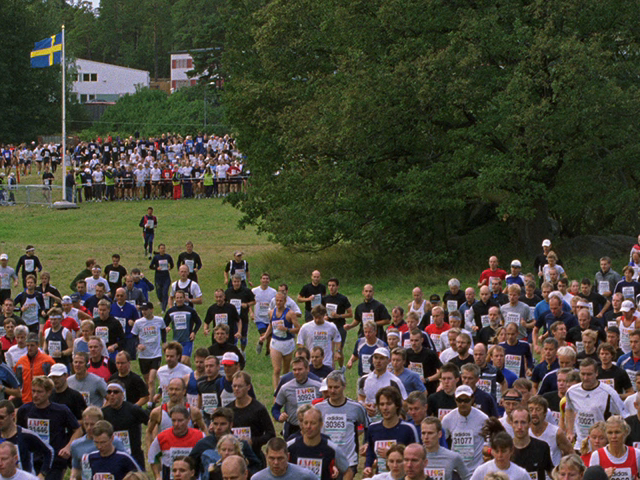} &
    \includegraphics[width=\imgwid]{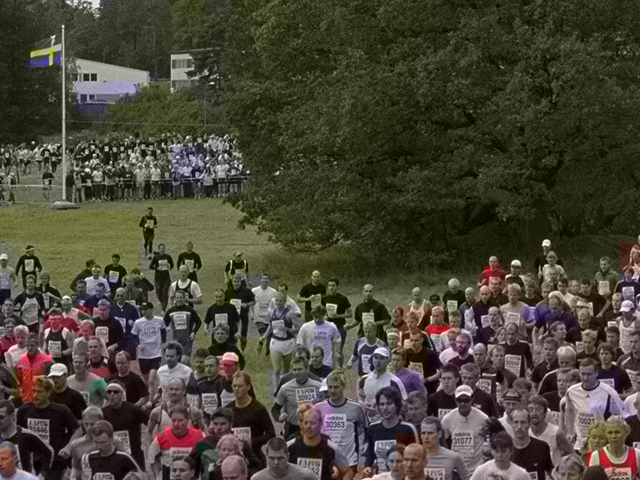} \\
    (a) $(\text{QP}_\text{Y},\text{QP}_\text{c})=(15,15)$ & 
    (b) $(\text{QP}_\text{Y},\text{QP}_\text{c})=(15,51)$ \\
    \includegraphics[width=\imgwid]{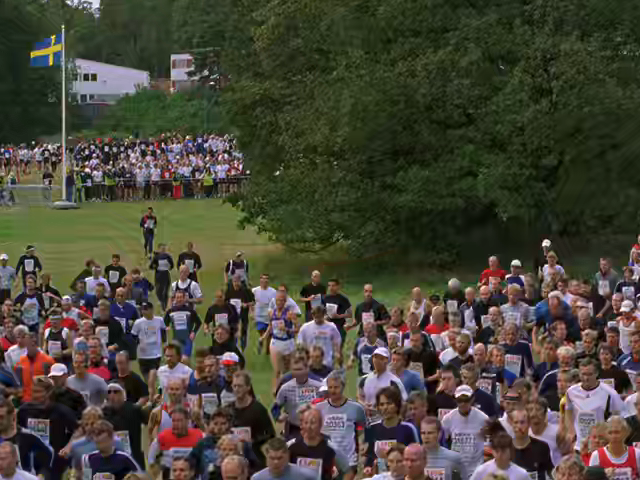} &
    \includegraphics[width=\imgwid]{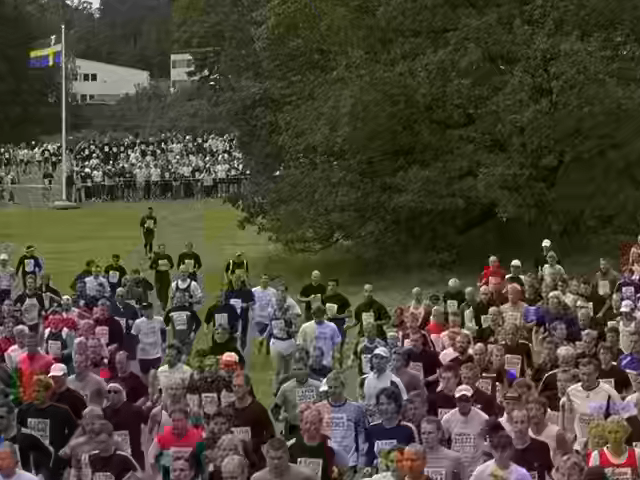} \\
    (c) $(\text{QP}_\text{Y},\text{QP}_\text{c})=(35,33)$ & 
    (d) $(\text{QP}_\text{Y},\text{QP}_\text{c})=(35,51)$
	\end{tabular}
	\caption{Exemplar frames encoded using different QP values on luma and chroma. (a) and (c) are encoded without increasing chroma QP values; (b) and (d) are encoded with extreme chroma QP's.}
	\label{fig:cqp_distortion_demo}
\end{figure}

As a preliminary step, we experimented with a ``stitching" method that generates two videos encoded at different QP values. These are decoded, then the luma of the first is combined with the chroma from the second. For example, one can combine a luma component encoded with $\text{QP}=15$, with chroma components encoded with $\text{QP}=51$ to create extreme chroma distortion. While this approach is intuitive and easy, it comes with a number of disadvantages. First, it does not produce bitstreams for every distorted video, hence RD performance is difficult or impossible to analyze. Moreover, we noticed that distortions created by this approach can appear very different from a real encoding result. Despite having the same quantization in chroma channels, their predictor/residual are considerably different. To address these shortcomings, we decided to create true encoding results by removing the clipping function (\ref{eq:transQP}) defined in the HEVC standard. An example of distortion generated in this way can be found in Fig. \ref{fig:cqp_distortion_demo}. As may be observed, both Figs. \ref{fig:cqp_distortion_demo}(b) and \ref{fig:cqp_distortion_demo}(d) show severe color shiftings on the runners. They are not exactly the same because of the different encoding outcomes from using different QP$_\text{Y}$ values. These unpleasant artifacts cannot be created without modifying the video codec standards.

\begin{figure}[!t]
	\centering
	\footnotesize
	\renewcommand{\tabcolsep}{1pt} 
	\renewcommand{\arraystretch}{0.6} 
	\def\imgwid{0.24\textwidth}
	
	\begin{tabular}{cc}
    \includegraphics[width=\imgwid]{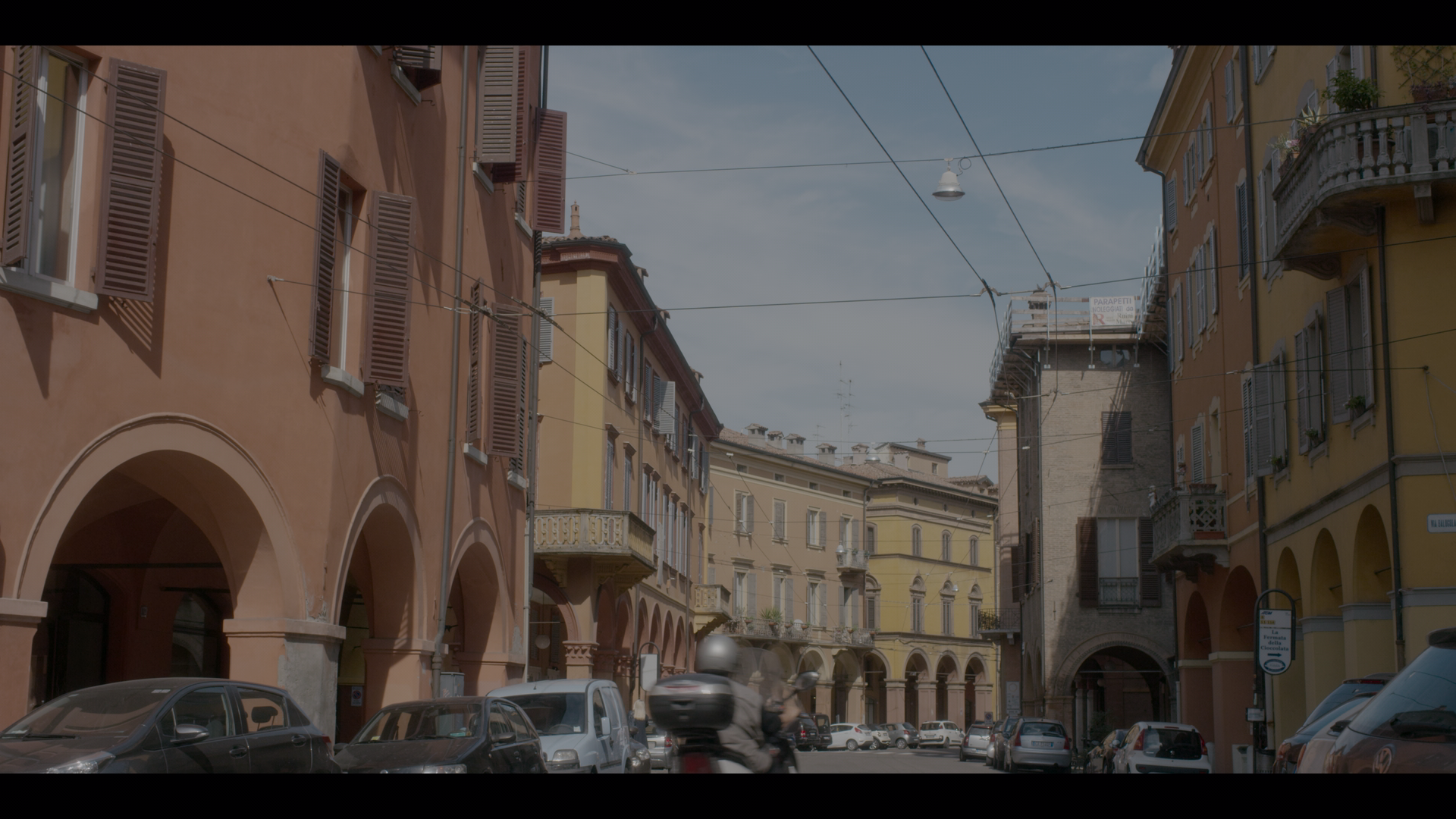} & 
    \includegraphics[width=\imgwid]{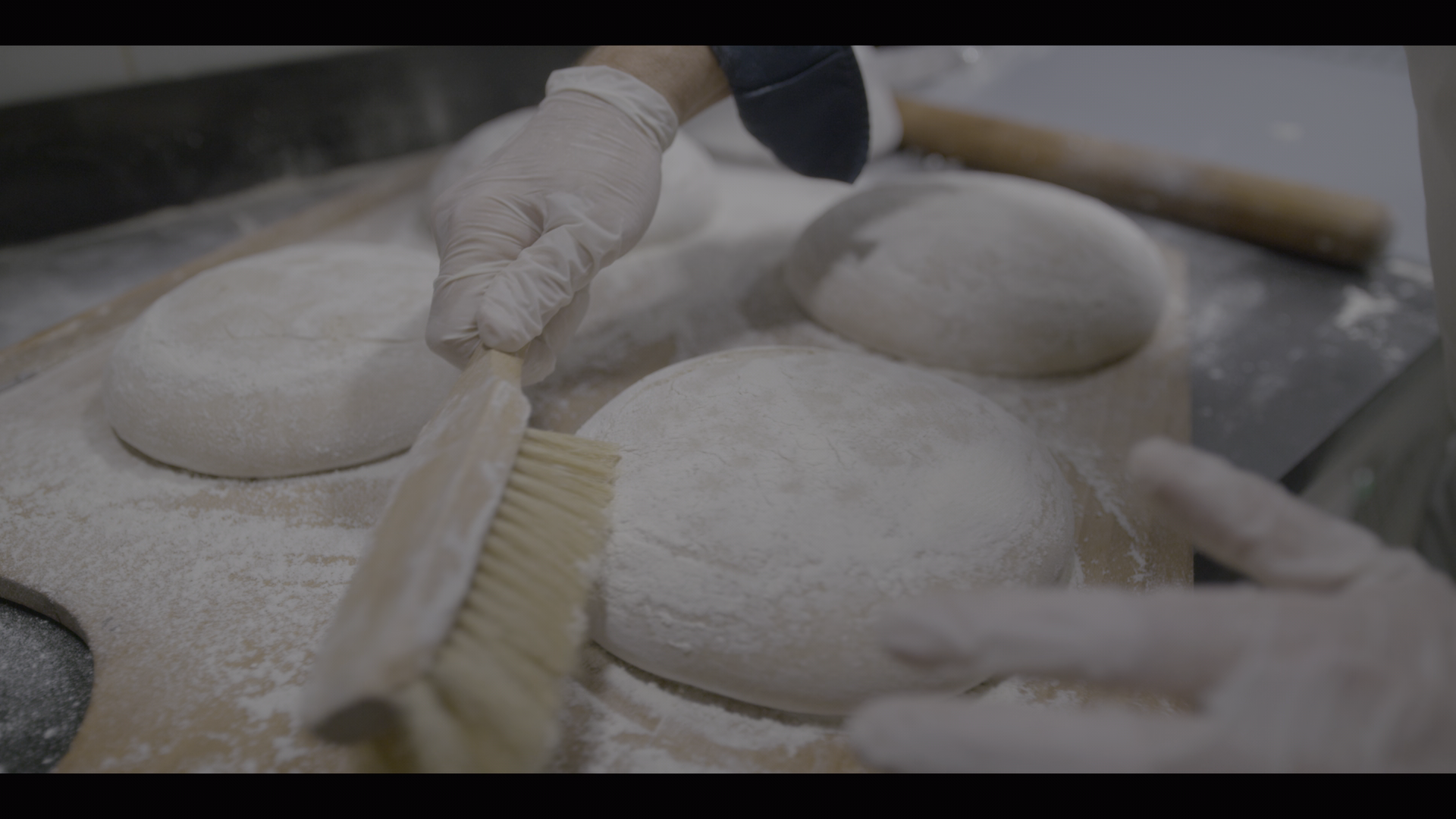} \\
    \includegraphics[width=\imgwid]{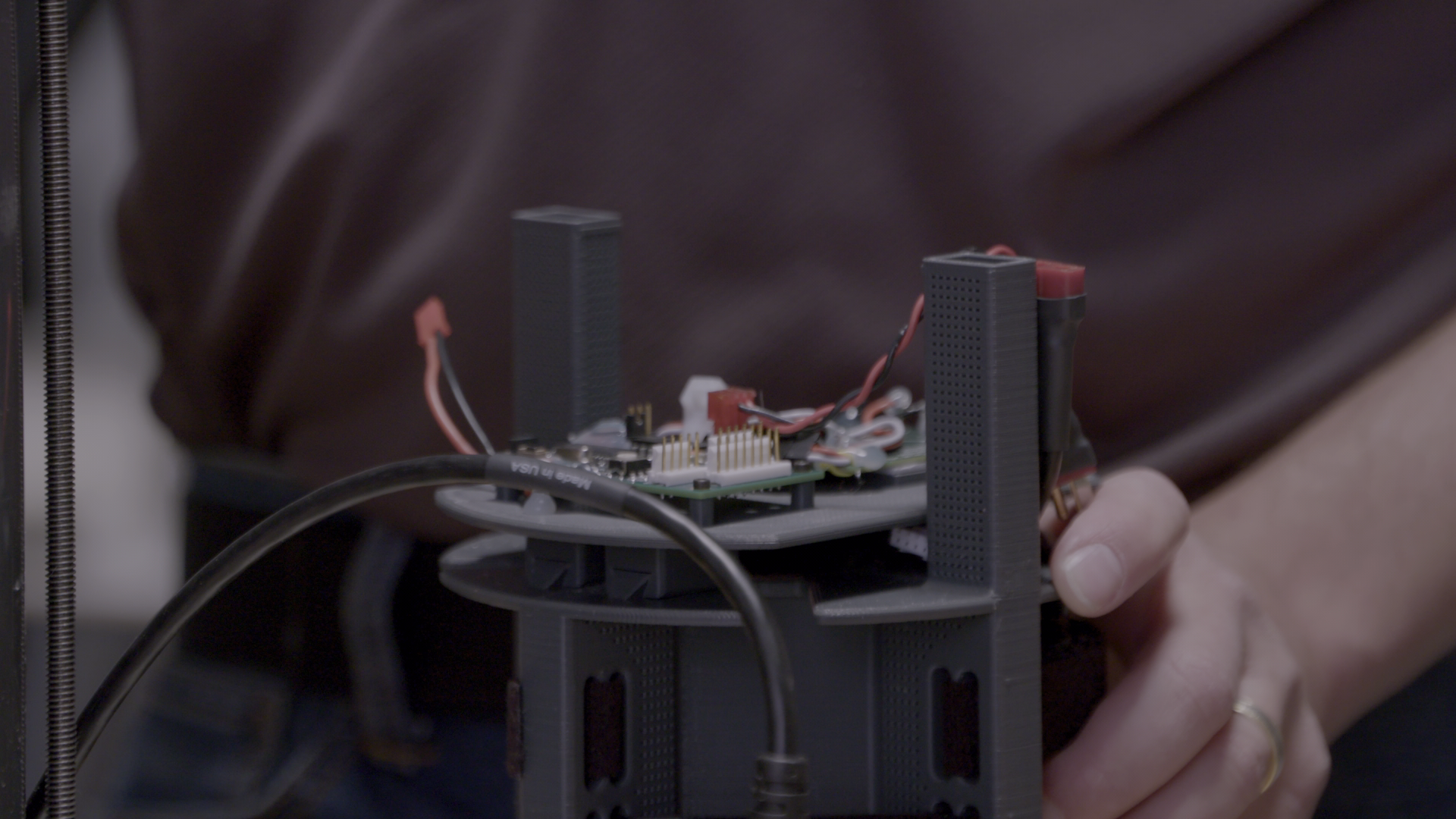} & 
    \includegraphics[width=\imgwid]{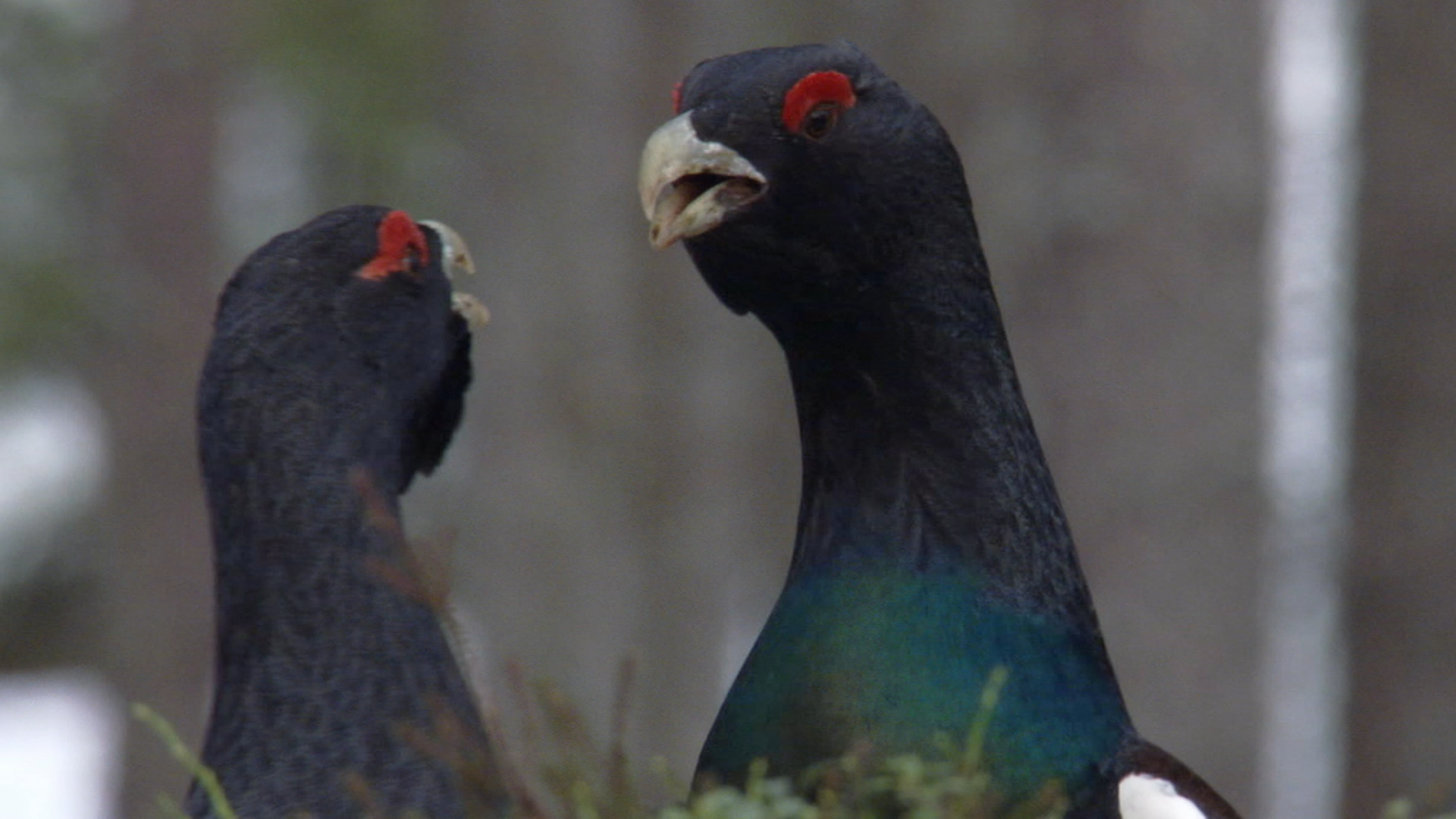} \\
	\end{tabular}
	\caption{Sample video frames extracted from the 21 video contents in the new database.}
	\label{fig:contents_src}
\end{figure}

\subsection{Source Sequences}
We selected 21 pristine Full High Definition (FHD) video sources from amongst the Netflix movies and TV shows in the Netflix catalog. These video frames were stored using visually lossless JPEG2000 compression. When needed, they were decoded to YUV420 format (YCbCr color space with 4:2:0 sampling). None of the videos contain audio components, which were removed. 
Fig. \ref{fig:contents_src} shows some of the selected content. The source videos have a variety of characteristics, such as dark/bright scenes, static/action scenes, close ups of human faces, and so on. All videos are about 8--10 second long and all of them have a frame rate of 24 frames per second.

\begin{figure}[!t]
	\centering
	\footnotesize
	\renewcommand{\tabcolsep}{0pt} 
	\def\imgwid{0.163\textwidth}
	
	\begin{tabular}{ccc}
    \includegraphics[width=\imgwid]{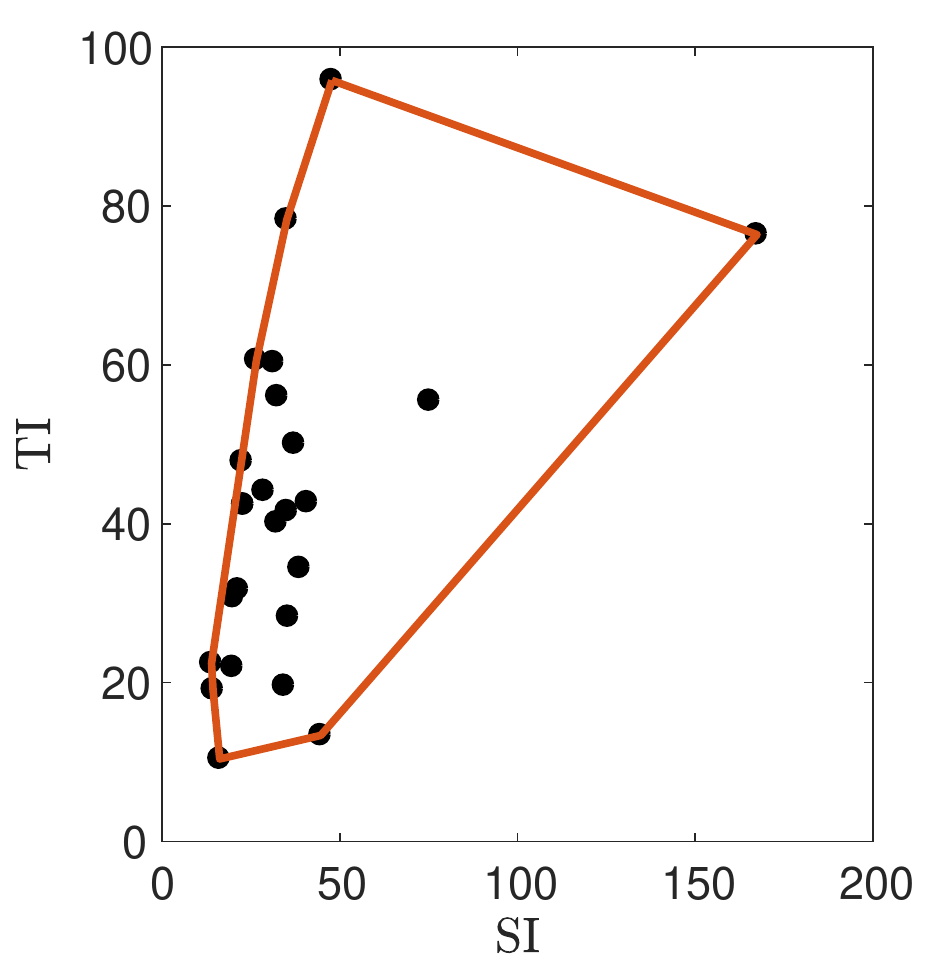} & 
    \includegraphics[width=\imgwid]{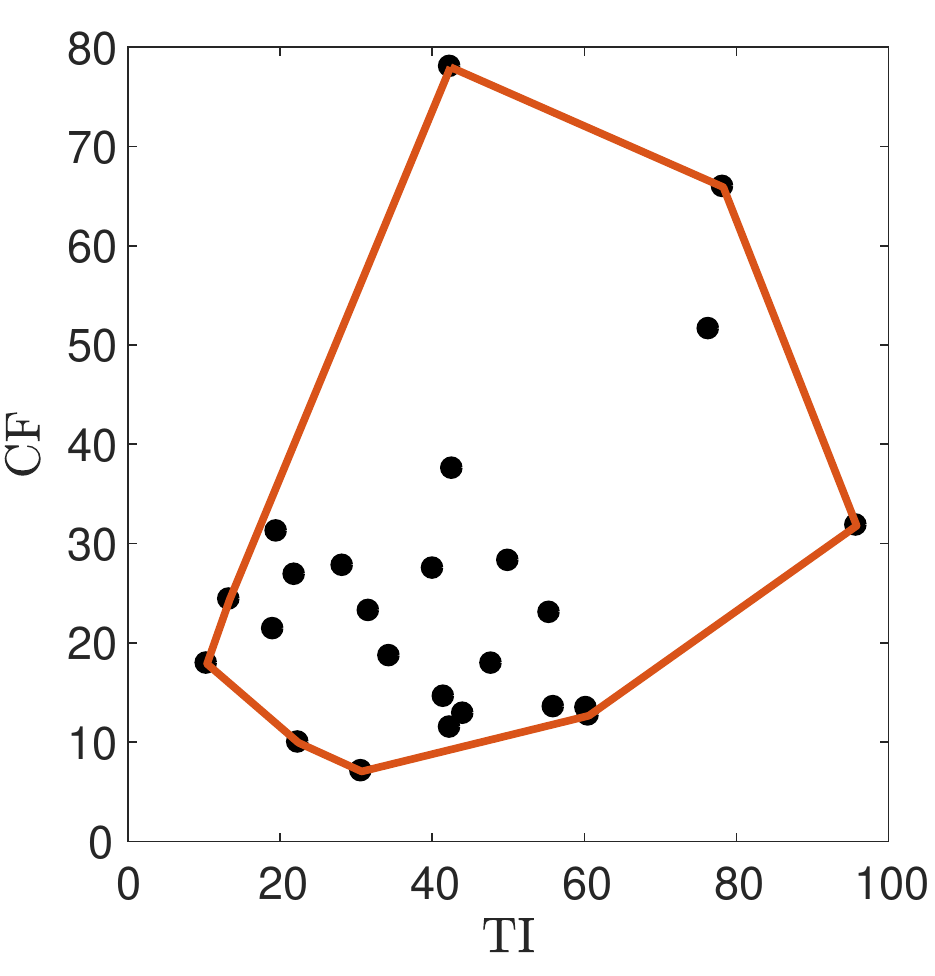} & 
    \includegraphics[width=\imgwid]{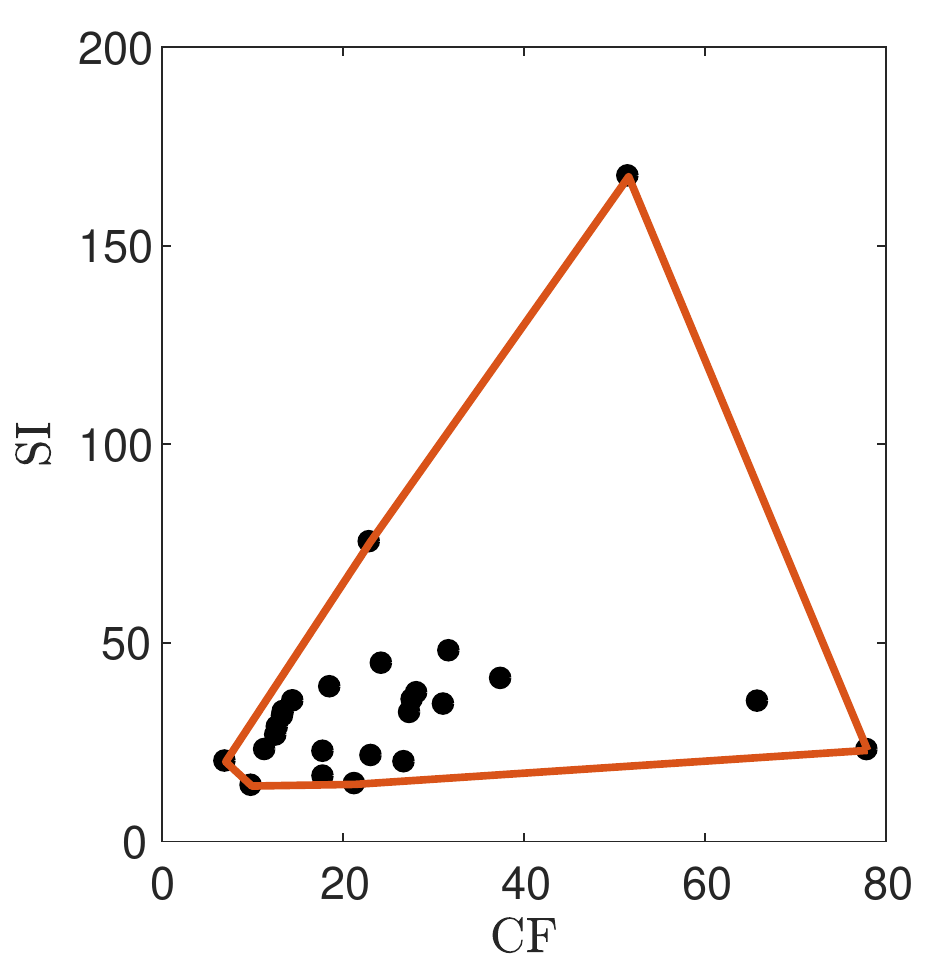} \\
	\end{tabular}
	\caption{Scatter plots and corresponding convex hulls of Spatial Information (SI), Temporal Information (TI), and Colorfulness (CF) of the video contents in our database.}
	\label{fig:si_ti_cf}
\end{figure}

As a way of quantifying the low-level content of the videos, we adopted the popular Spatial Information (SI), Temporal Information (TI), and Colorfulness (CF) indices \cite{GhadiyaramStalling, winkler2012analysis, Buczkowski, MXChen2020}. To obtain SI, the luma channel is processed by horizontal and vertical Sobel filters to obtain responses $s_h$ and $s_v$. Let $s_r = \sqrt{s_h^2+s_h^2}$ denote the edge magnitude at each pixel location. The SI value is then simply the maximum standard deviation of $s_r$
\begin{equation}
    \text{SI} = \max_t\left(\sigma_{s_r}\right),
\end{equation}
where $\sigma_{s_r}$ are calculated on a per-frame basis, and $t$ denotes frame index. The TI value is computed as the standard deviation of the differences between adjacent frame pairs
\begin{equation}
    \text{TI} = \max_t\left(\sigma_{\Delta I_t}\right),
\end{equation}
where $\Delta I_t = I_t - I_{t-1}$ is the pixel-wise difference between the $t^\text{th}$ frame and the $\left(t-1\right)^\text{th}$ frame. Finally, color information is captured using CF index in \cite{Hasler2003}, which is formulated as
\begin{equation}
    \text{CF} = \max_t\left(\sqrt{\sigma_{rg}^2+\sigma_{yb}^2} + 0.3\sqrt{\mu_{rg}^2+\mu_{yb}^2}\right),
\end{equation}
where $rg=R-G$ and $yb=0.5(R+G)-B$ represent opponent color spaces, while $\mu_x$ and $\sigma_x$ are the mean and standard deviation of a plane $x$. Before calculating CF, we transformed YUV420 to RGB444. As shown in Fig. \ref{fig:si_ti_cf}, the reference videos of our database widely span the SI-TI-CF space.

\begin{figure}[!t]
\centerline{
\includegraphics[width=0.92\columnwidth]{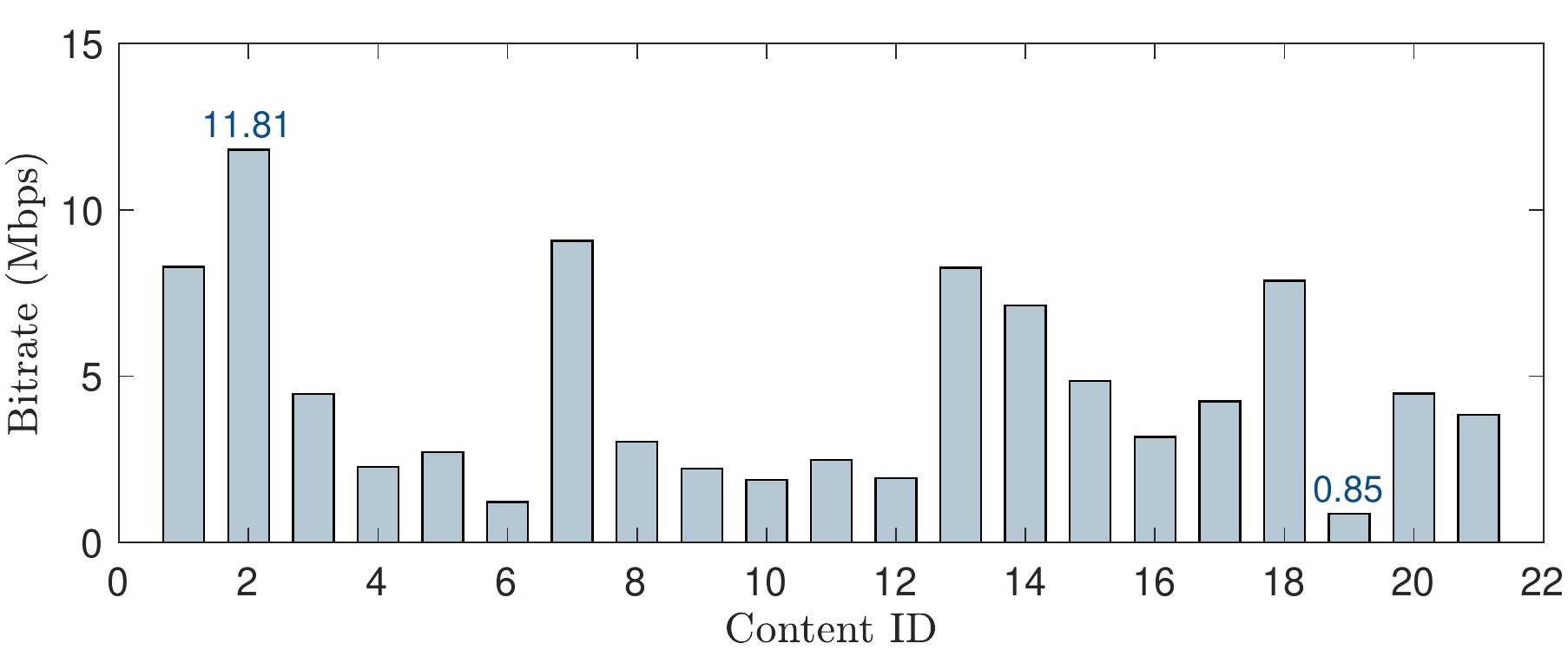}
}
\caption{Encoding complexity of each video content in the new database, expressed as the bitrate derived by encoding them using libx264 with a fixed CRF of 23.}
\label{fig:complexity}
\end{figure}

In addition to the typical SI-TI-CF plots, we used another strong indicator of video complexity. Specifically, we encoded all video contents using libx264\footnote[2]{https://trac.ffmpeg.org/wiki/Encode/H.264} at a fixed Constant Rate Factor (CRF) setting of 23. Then, the bitrate of each encoded video was measured as an empirical indication of encoding complexity \cite{bampis2018spatiotemporal} (intuitively, complex contents consume more bits than simple contents at the same quantization level). Figure \ref{fig:complexity} demonstrates that our video contents span a large range of the bitrate spectrum at constant CRF, ranging from less than 1.0 Mbps up to around 11.8 Mbps. Due to content licensing restrictions, the video contents of the database cannot be made available.


\subsection{Encoding Space and Experimental Design}
In subjective experiments of video quality, it is important to carefully design the distortion space, such that the video artifacts are perceptually well-separated and the number of video distortions is not too large. A large number of video distortions will either increase the average viewing time for participants or the number of required participants, both of which are undesirable in subjective lab tests.

\begin{figure}[!t]
  \centerline{
  \includegraphics[width=1.0\columnwidth]{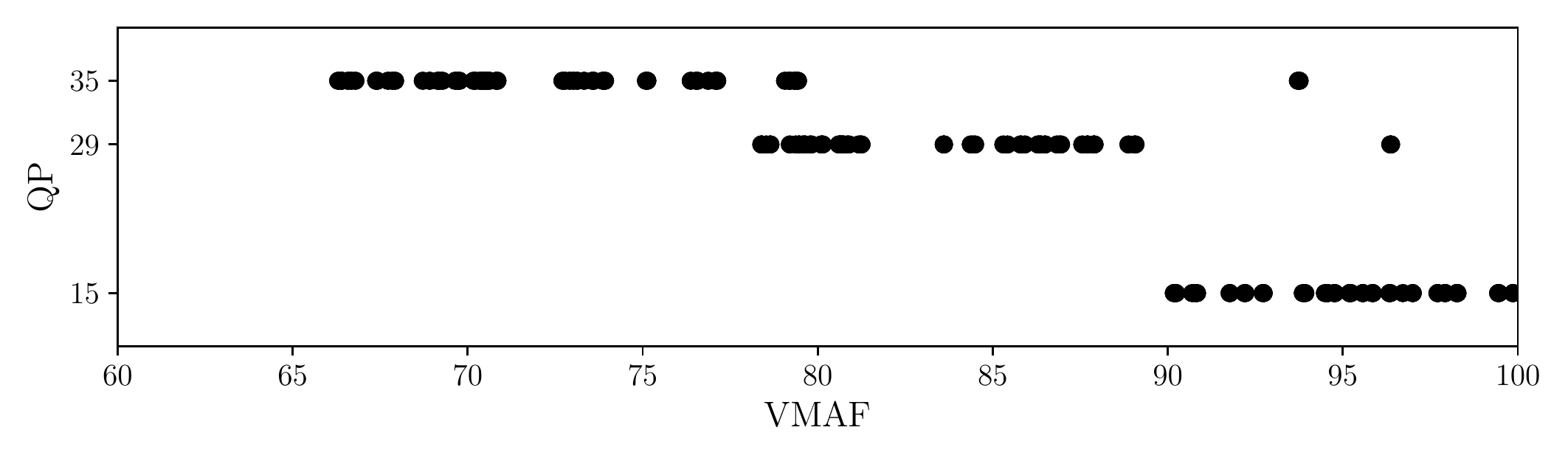}
  }
  \caption{VMAF scores of distorted videos against their luma QP. Each point represents a distorted video in the dataset.}
  \label{fig:vmafdist}
\end{figure}

We carefully selected three luma QP values such that their VMAF distributions were separated, as demonstrated in Fig. \ref{fig:vmafdist}. In fact, the QP values give visually separated distortion levels on most of contents. The quantization parameters we used are summarized in Table \ref{chroma_qp_selection}, where QP$_\text{Y}$ represents luma QP, and QP$_\text{c}$ represents chroma QP. For each QP$_\text{Y}$ employed on a content, three \textbf{cb\_qp\_offset\,/\,cr\_qp\_offset} settings, ranging from $0$ to $51$, were assigned resulting in three different QP$_\text{c}$s. Aside from the two extreme values, we uniformly covered the range of QP$_\text{c}$ for each content. For example, a content encoded with QP$_\text{Y}=15$ can bracket one set of QP$_\text{c}$ from the sets $\{15,21,51\}$, $\{15,27,51\}$, or $\{15,39,51\}$. After the QP selection process, the resulting distorted videos were stored for later display to human participants.

\begin{table}[t]
  \caption{Design of quantization distortion levels in the new database. For each content, one of the three values in the brackets was selected for the study.}
  \centering
  \scalebox{1.0}{
    \renewcommand{\arraystretch}{1.07} 
  \begin{tabular}{|c|c|c|}
  \hline
  QP$_\text{Y}$ & \textbf{cb\_qp\_offset\,/\,cr\_qp\_offset} & QP$_\text{c}$ \\ \hline
  15 & 0, \{6, 12, 30\}, 51 & 15, \{21, 27, 39\}, 51\\ \hline
  29 & 0, \{6, 12, 20\}, 51 & 29, \{33, 36, 43\}, 51\\ \hline
  35 & 0, \{6, 12, 17\}, 51 & 33, \{36, 41, 46\}, 51\\ \hline
  \end{tabular}}
  \label{chroma_qp_selection}
\end{table}

Following standard practice, the subjective study sessions were divided into several separate 30-minute viewing sessions, to avoid users' and/or visual fatigue. In addition, every subject's sessions were separated by at least 24 hours, for the same reason \cite{ITUTBT50011}. Hence, we designed a two-session study as follows. The videos were assumed to have durations of 10 seconds each (actually 8--10s), and both sessions presented the same 21 contents.

In the first session for each content, five distorted versions of it and one pristine video were included for each content. Therefore, the corresponding viewing time of the first session was about 
\begin{equation}
 21.0 \text{ min} = 10 \text{ sec} \times
 \overbrace{
 21
 \times \underbrace{(5+1)}_{\mathclap{\text{distorted + pristine}}}
 }^{\mathclap{\text{\# of videos}}}.
\end{equation}
Similarly, the viewing time of the second session was about
\begin{equation} 
 17.5 \text{ min} = 10 \text{ sec} \times 21 \times (4+1).
\end{equation}
To allow for variations, 30 minutes were allocated to each session, including the instructions given each subject in the first session.

\subsection{Viewing Conditions}
During the study, all the videos and graphical user interface (GUI) were displayed using a 15 inch MacBook Pro having a \SI[mode=text]{0.391}{\metre} diagonal length and a $2880\times1800$ native resolution retina display. During the subjective test, we set the display resolution to $1920\times1200$. The display was positioned at a viewing distance of about \SI[mode=text]{0.610}{\metre} (2 feet), which is approximately equivalent to three times the height of the display monitor. Also, the subjects were told not to modify their viewing positions very much, while still remaining comfortable.

We set up a controlled laboratory environment for the subjective study, obeying the recommended viewing conditions for subjective assessments in a \textbf{laboratory environment} in section 2.1.1 of BT.500-11 \cite{ITUTBT50011}. Specifically:
\begin{enumerate}
\item The ambient illumination was fixed at a low light level. The brightness level of the display was held constant at 50\% of maximum throughout the study, and the automatic brightness adjustment feature was disabled to guarantee consistency of brightness.
\item We measured and calibrated the lighting conditions using a Sekonic L-758CINE photometer before each session. In the environment that we set up, the luminance of the background wall was about \SI[mode=text]{0.8}{\candela/\metre^2}. We measured the brightest luminance on the screen from displayed white level, which was \SI[mode=text]{5}{\candela/\metre^2}, at a distance of about 2 feet (the location of the subject) in the darkened room.
\item Accordingly, the ratio of luminance of background behind picture monitor to peak luminance of picture was given by
\begin{equation} 
R_{bm} = \frac{0.8~\si{\candela/\metre^2}}{5~\si{\candela/\metre^2}}=0.16,
\end{equation}
which is very close to the recommended ratio of $0.15$.
\end{enumerate}

\begin{figure}[!t]
  \centerline{
  \includegraphics[width=0.92\columnwidth]{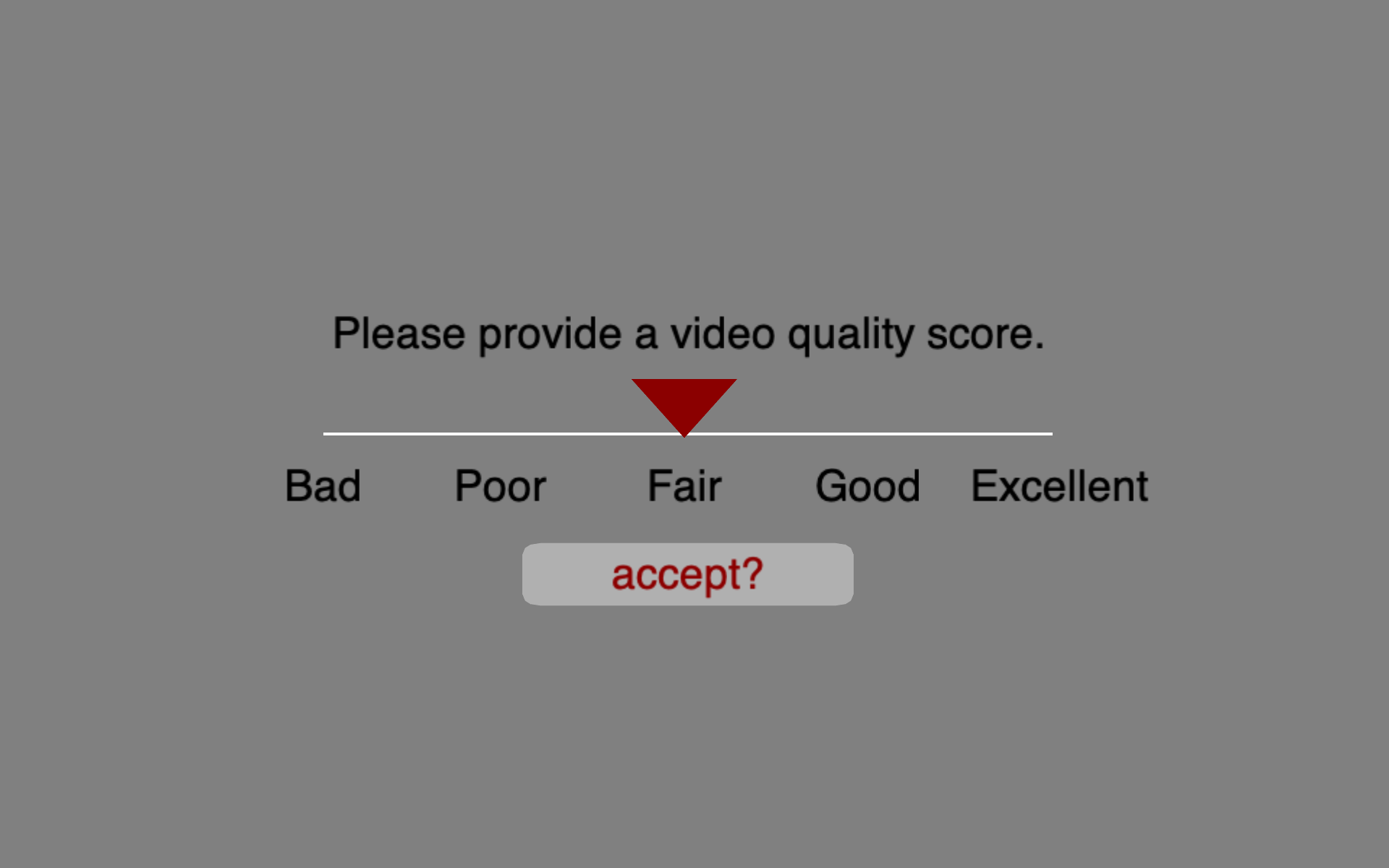}
  }
  \caption{User interface employed in the subjective test: Likert rating bar for the subject to submit a quality score for the video they completed viewing.}
  \label{fig_uinterface}
  \end{figure}

\subsection{Study Instructions, Training, and Subjects}
Each participant was asked to take a Snellen visual acuity test and an Ishihara color test, to understand their state of vision. If a subject normally used corrective lenses when watching videos, they were asked to wear them to achieve normal vision when participating in the study. Moreover, a set of instructions were given to each subject explaining the subjective testing process. They were asked to report opinion scores expressive of their viewing experience on every video. The detailed written instructions were given, followed by a brief verbal exchange to ensure that the subjects understood their tasks. In order to remove any possible rating biases, the participants were requested not to base their judgments on whether or not any video content was interesting. Finally, the subjects were informed that there were no right or wrong answers in the experiment. 

\begin{figure}[tp]
  \centerline{
  \includegraphics[width=0.805\columnwidth]{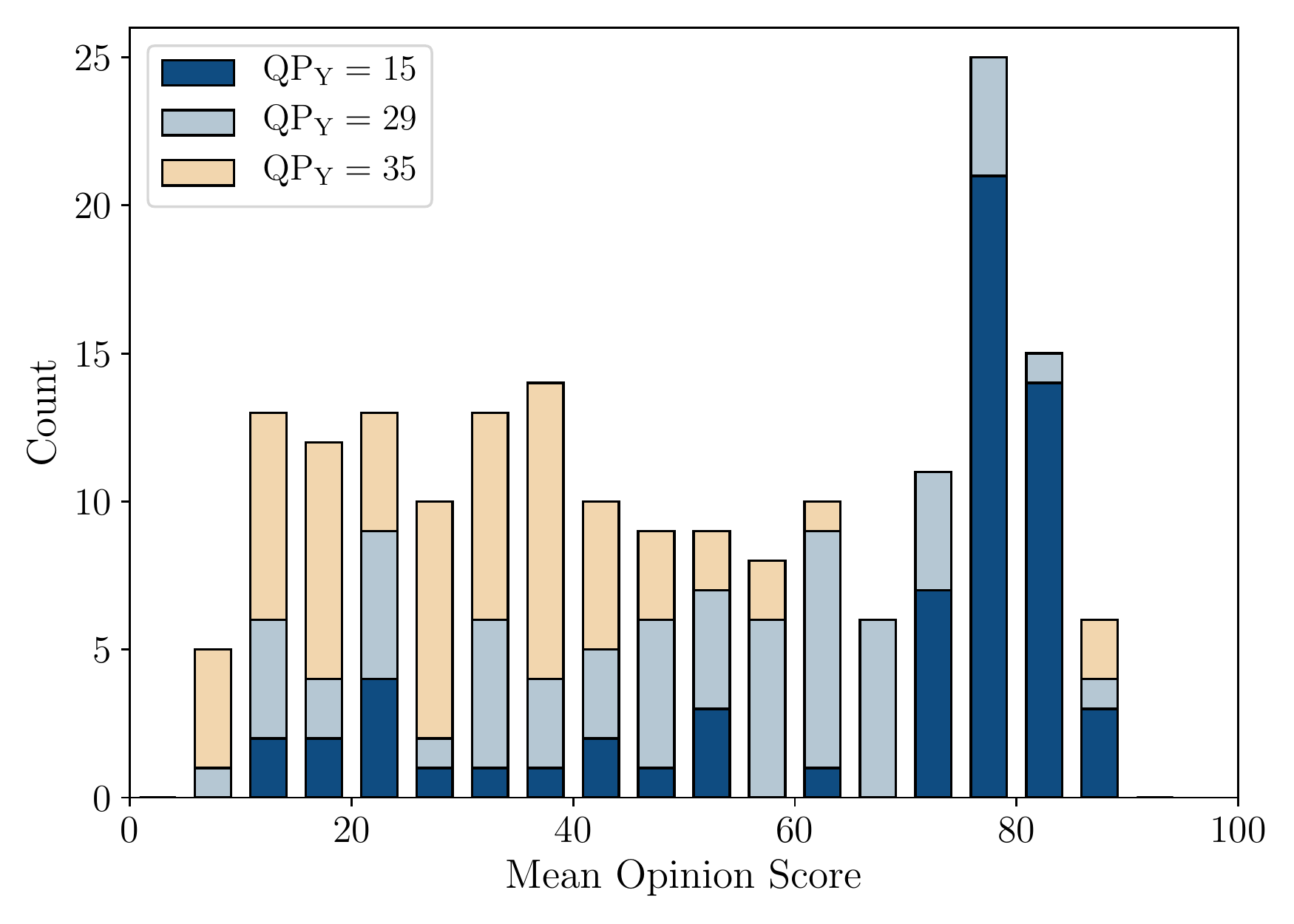}
  }
  \caption{MOS distribution across the NFLX-Color database. A further breakdown with respect to QP$_\text{Y}$ were shown in each bin.}
  \label{fig:mos_hist}
\end{figure}

As a practice, four ``practice" videos were presented before the actual study began. These videos were designed to be broadly representative of the range of distortion types and levels that the participants might experience during the actual test. This \textbf{``training session"} facilitated the subject being familiar with the study, and was only given in the first session. The contents used for training were the same for all subjects and were distinct from the actual test contents. The \textbf{``testing session"} will be described in the next subsection.

The study was carried out over a three-week period at Netflix. In total, 34 subjects were voluntarily recruited. Roughly half of the subjects were experts in the field of image/video engineering while the others were average viewers. In terms of visual acuity, 13 subjects (38.2\%) possessed approximately 20/25 vision (slightly worse than ideal) while the others had 20/20 or better vision after correction. Only two participants did not accurately recognize Ishihara Plates during the color vision test, but were allowed to continue as being typical of the populace. Interestingly, the two subjects who failed the color vision test were able to perceive the chroma distortions, and were not rejected as outliers.

\begin{figure*}[!ht]
	\centering
	\footnotesize
	\renewcommand{\tabcolsep}{0pt} 
	\def\imgwid{0.295\textwidth}
	
	\begin{tabular}{ccc}
    \includegraphics[width=\imgwid]{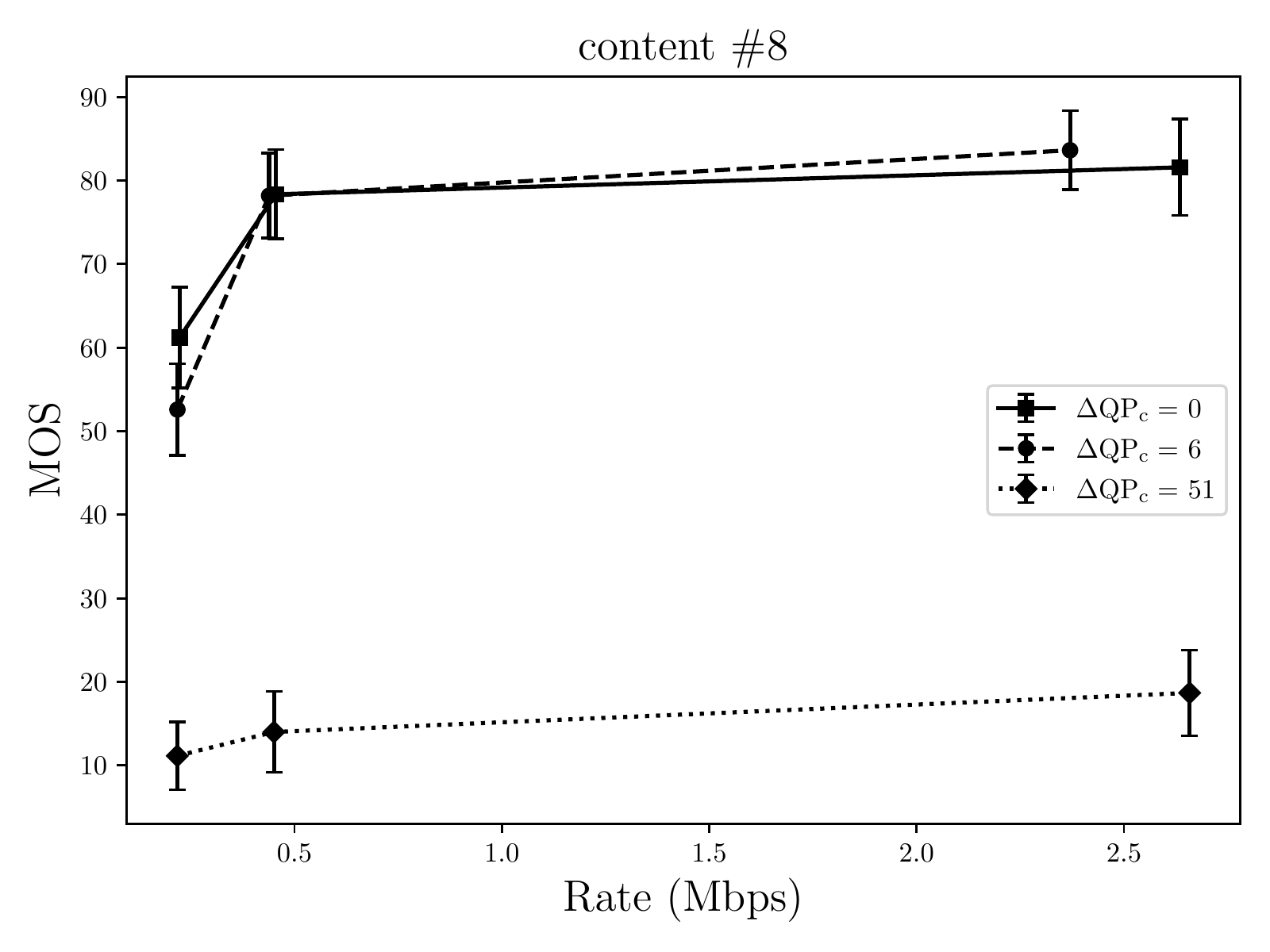} & 
    \includegraphics[width=\imgwid]{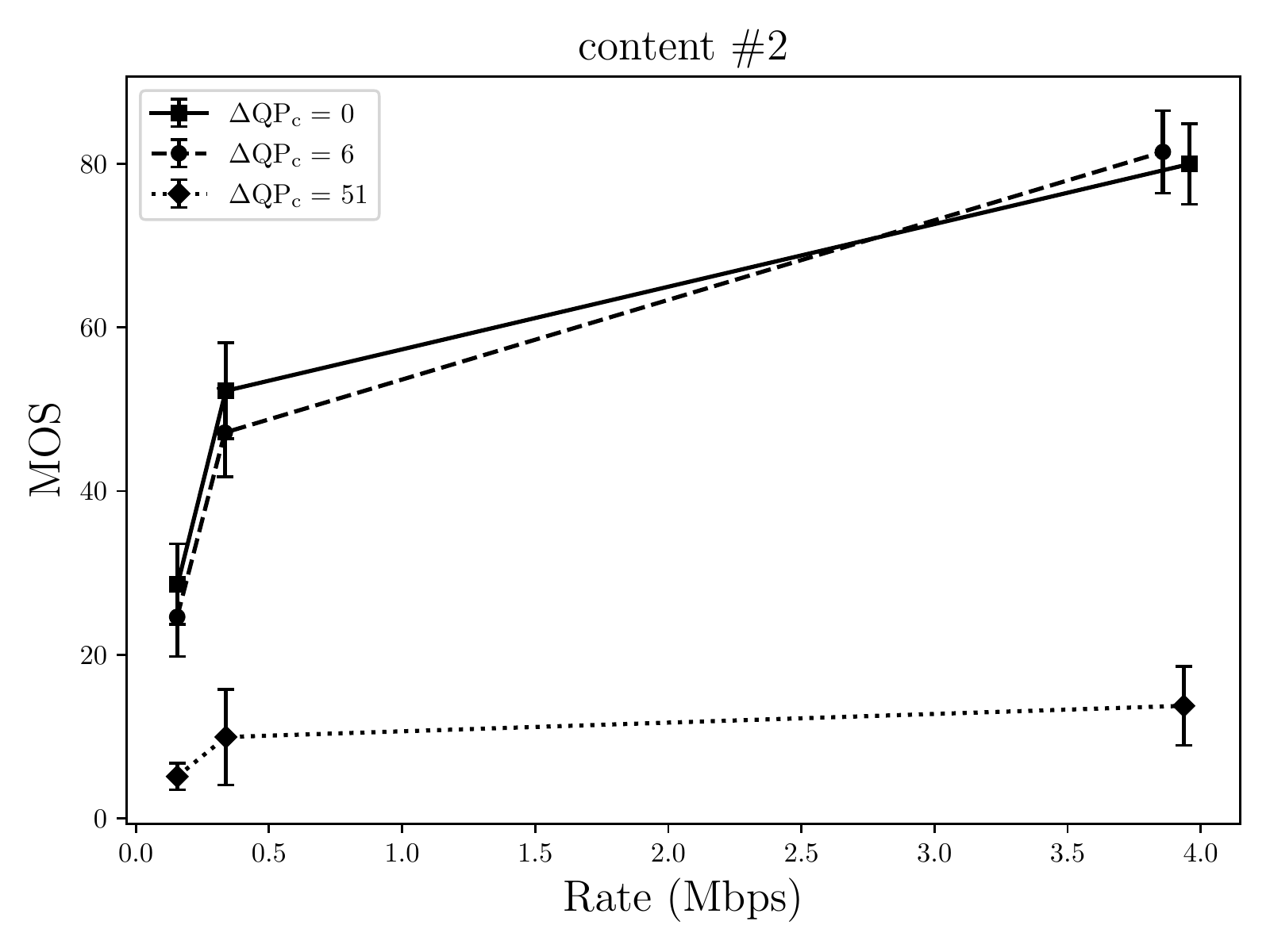} & 
    \includegraphics[width=\imgwid]{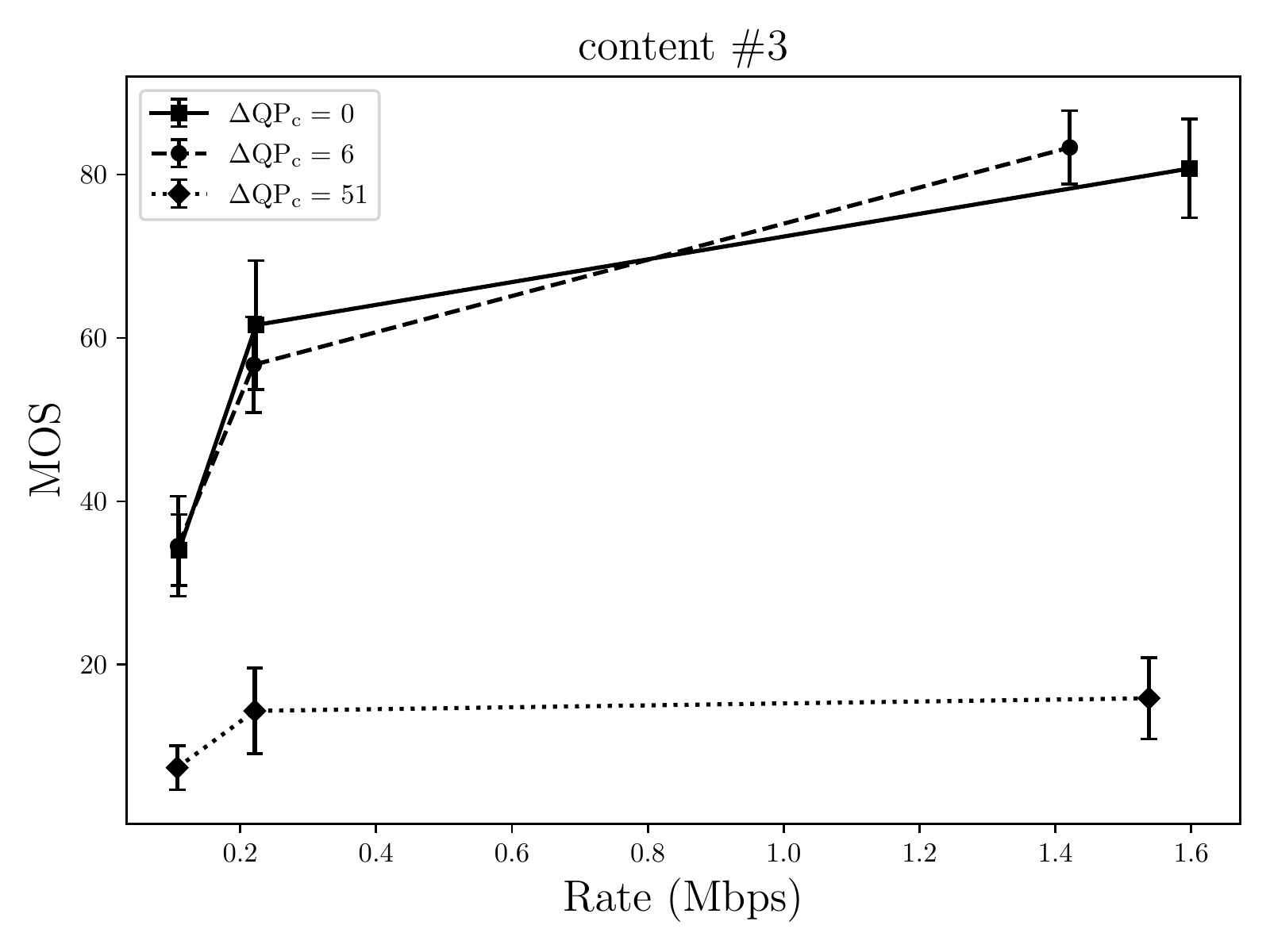} \\
    \includegraphics[width=\imgwid]{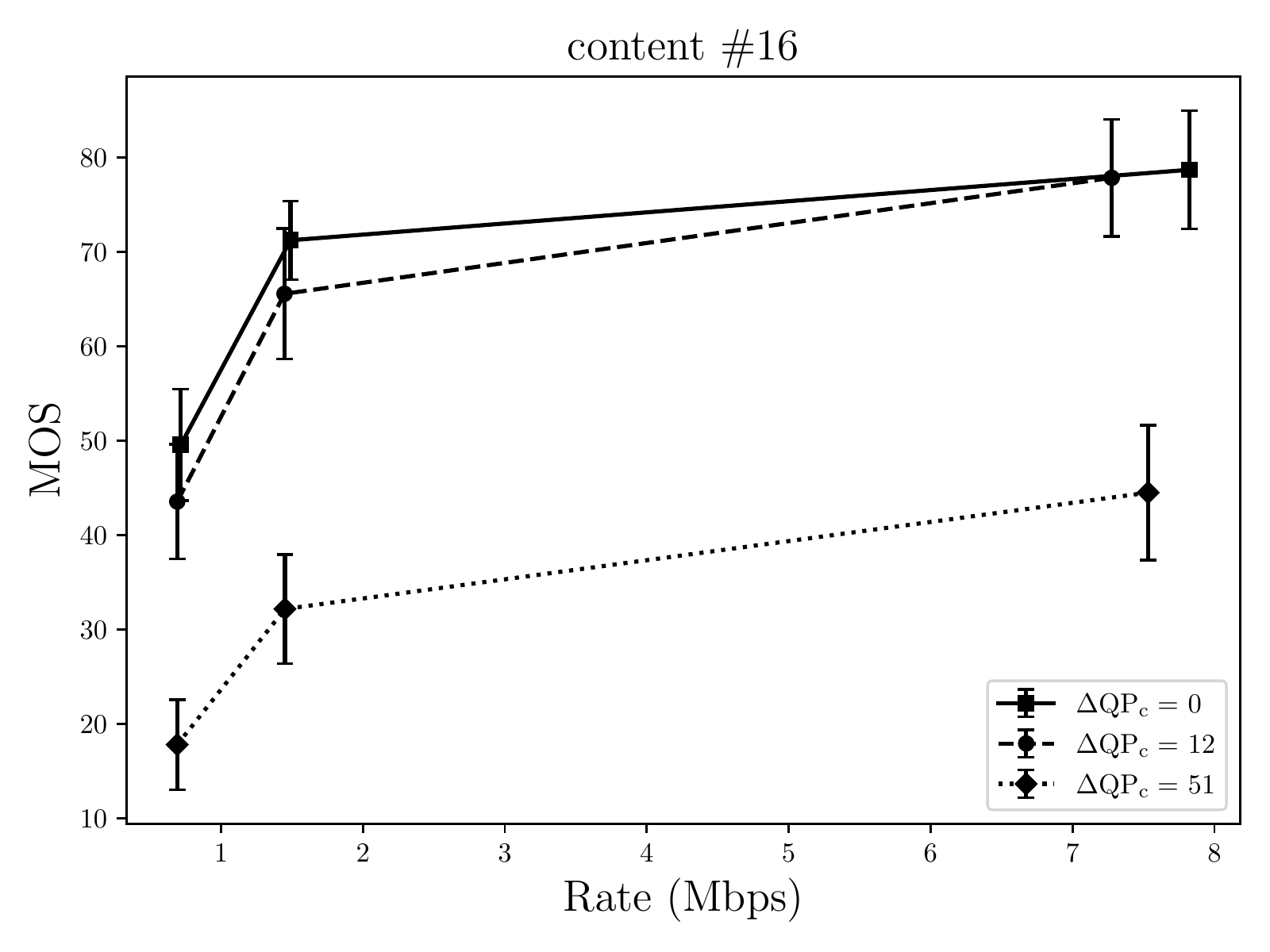} & 
    \includegraphics[width=\imgwid]{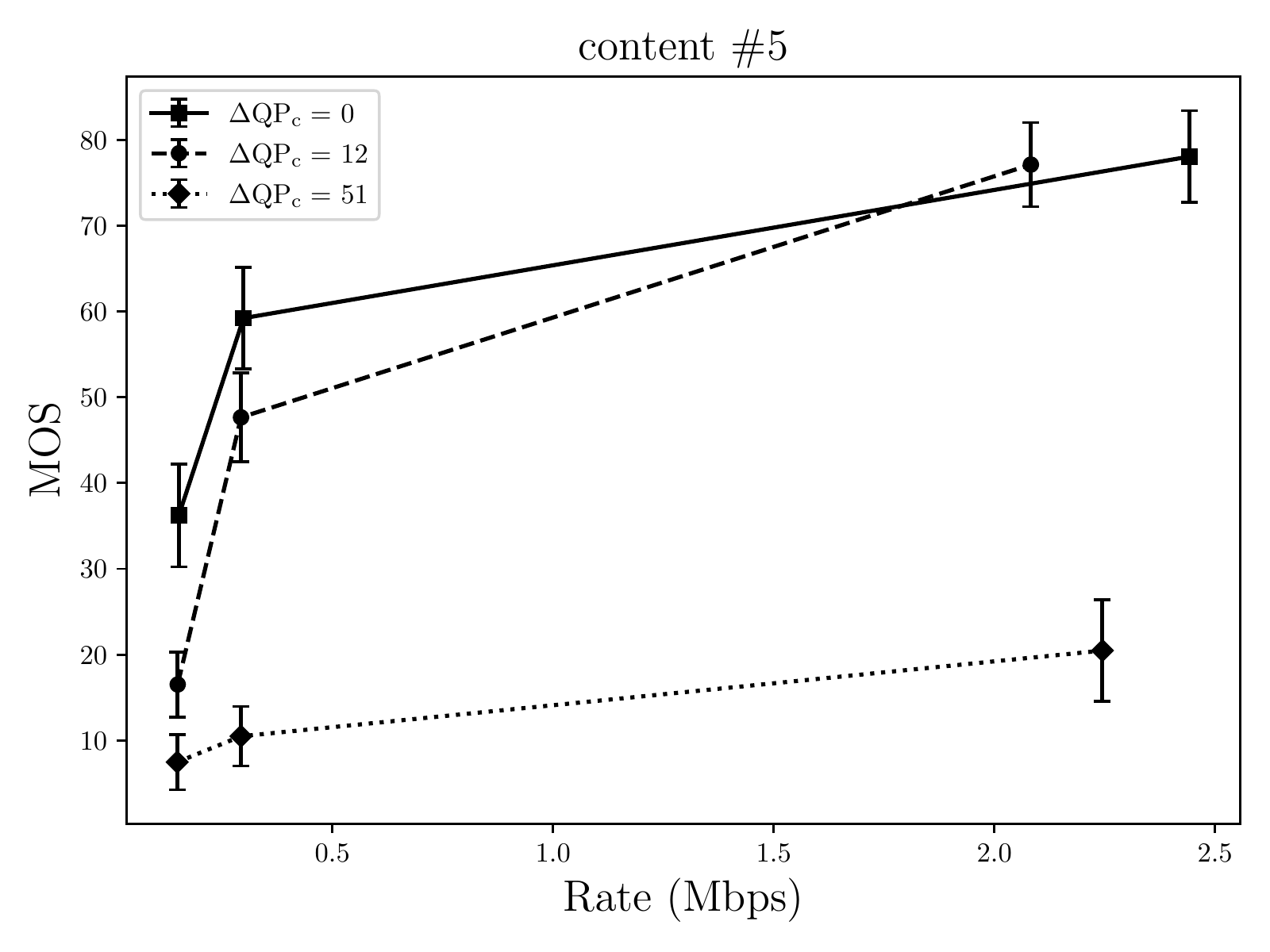} & 
    \includegraphics[width=\imgwid]{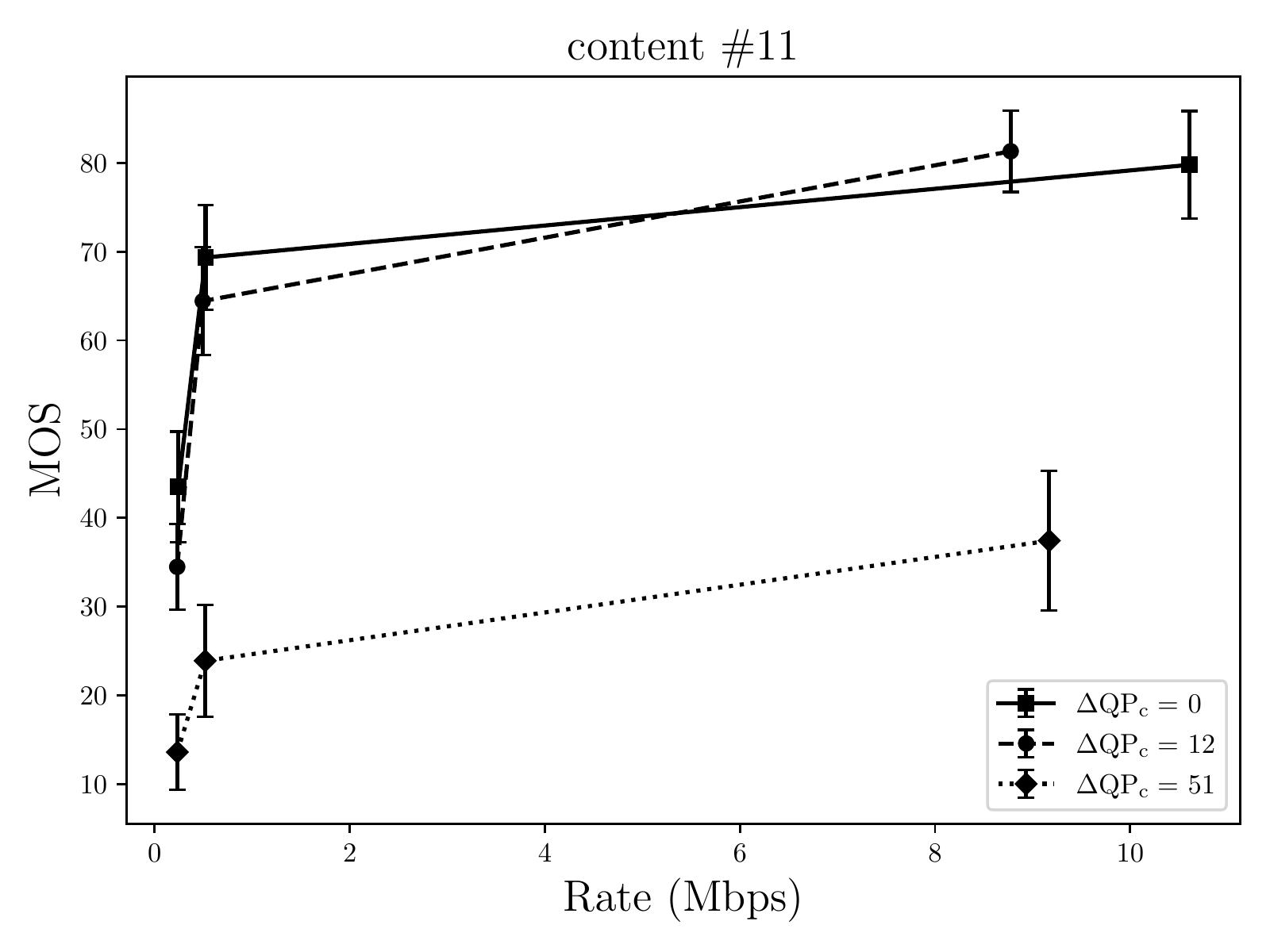} \\    
	\end{tabular}
	\caption{Examples of MOS Rate-Distortion curves for different \textit{chroma\_qp\_offset} (denoted by $\Delta \mathrm{QP_c}$) settings. The error bars indicate the 95\% confidence interval of each RD point.}
	\label{fig:mos_rd}
\end{figure*}

\subsection{User Interface and Test Methodology}
The user interface was designed using PsychoPy3 \cite{Peirce2007}, which is an open source software often used in experimental psychology and neuroscience research. We followed the single stimulus procedure described in \cite{itup910}, whereby the videos were displayed one after the other. 
Each video was displayed at native resolution to prevent additional scaling artifacts. Two $1920\times60$ black bars were rendered at the top and the bottom of the screen, respectively, to fill the 16:10 aspect ratio of the MacBook Pro display. 

At the end of each displayed video, a continuous Likert scale of possible video quality scores was displayed on the screen, as demonstrated in Fig.~\ref{fig_uinterface}.
The cursor was reset to the center of the rating bar after each rating, to avoid bias. Five equally spaced labels ``Bad," ``Poor," ``Fair," ``Good," and ``Excellent," were marked below the rating bar to help the subject understand the range and types of ratings they could apply. This Likert scale is similar to the ACR scale documented in ITU-R \cite{ITUTBT50011}. To rate the videos, the subjects moved the red inverted triangle cursor above the rating bar using the touchpad on the MacBook Pro. After the subject decided a quality score that corresponded to his or her perception, the subject was asked to click on the \textit{Accept} button to record the score and to view the next video. After submitting the score, the position of the sliding bar was converted to an integer quality score in the range $\left[1,100\right]$. Then, the next video clip was presented. It should be noted that once a score was submitted, the name and score of that video were written to file and could not be changed. Of course, the video could not be viewed again.

In this way, we collected 6468 scores over 56 sessions from 34 subjects. In the following, we will refer to the videos and the data collected during the subjective study as the NFLX-Color (NFLX$_\text{c}$) video quality database.

\section{Analysis of the Subjective Experiment}
The following subsections present an analysis of the subjective experiment results.
\subsection{Data Processing}
Given the raw scores collected from the study, subjective Mean Opinion Scores (MOS) were then computed according to the procedures detailed in \cite{Seshadrinathan2010}. Let $s_{ijk}$ denote the raw score assigned by the $i$-th subject on the video $j$ in session $k$, where $k \in \{1, 2\}$. The raw scores were then converted into z-scores for each session:
\begin{equation} 
  z_{ijk} = \frac{s_{ijk}-\mu_{ik}}{\sigma_{ik}},
\end{equation}
where $\mu_{ik}$ and $\sigma_{ik}$ are the mean and the standard deviation of the raw scores over all videos assessed by subject $i$ in the $k^\text{th}$ session. Then, the subject rejection procedure described in ITU-R BT500.13 \cite{ITUTBT50013} was conducted to remove outliers. After this procedure, we linearly scaled the remaining z-scores to $[0, 100]$. The MOS of each video was the mean value of the rescaled z-scores from the remaining subjects. In our study, we used the SUREAL\footnote[3]{https://github.com/Netflix/sureal} python package to calculate MOS with subject rejection following the procedure described above, and only 6 out of 34 subjects were rejected. 

\subsection{Analysis of Subjective Quality Scores}
Figure \ref{fig:mos_hist} plots the distribution of MOS obtained by the aforementioned procedures. Clearly, the MOS of the distorted videos span the entire range of quality. A peak may be observed at a high MOS region, since slightly increasing the chroma QP (e.g. $\Delta\text{QP}_\text{c}=6,12$) on the videos with $\text{QP}_\text{Y}=15$ did not affect subjective quality much. We also computed the standard error of the subjective scores on each video, obtaining an standard error ranging from 0.822 to 4.414, with an average value of 2.877. This indicates the difficulty of the quality prediction task on specific contents or distortions. We found bifurcated opinions from the subjects occurred on combinations of a ``clean” luma plane and severely distorted chroma channels. This is understandable: the recipe $(\text{QP}_\text{Y}, \text{QP}_\text{c})=(15,51)$ is an encoding setting that does not usually appear in daily life; hence they received more inconsistent scores. On the contrary, consensus emerged on the other corner, where both luma and chroma were distorted at the highest quantization level. 

\begin{table*}[!ht]
  \renewcommand{\arraystretch}{1.3}
  \caption{Performance comparison of candidate chromatic features: Each cell shows the SROCC values of a feature applied on $\mathrm{Cb}$ and $\mathrm{Cr}$ channels, expressed as SROCC$\mathrm{_{Cb}}$\,/\,SROCC$\mathrm{_{Cr}}$. The best results among each dataset are denoted by boldface. Please refer to the dataset acronyms in section \ref{sec:dataset}.}
    \label{tab:chroma_feature_srcc}
    \centering
    \renewcommand{\tabcolsep}{3.8pt} 
    \begin{tabular}{l c c c c c c c c c c}
    \hline\hline
    Dataset & 
    LIVE-VQA & LIVE-MBL & NFLX & \textbf{NFLX}$\mathbf{_c}$ & BVI-HD & CSIQ-VQA & EPFL        & VQEG      & SHVC  \\
    \hline
    PSNR &
    0.074\,/\,0.146 & 0.559\,/\,0.548 & 0.580\,/\,0.669 & 0.689\,/\,0.705 & 0.408\,/\,0.398 &
    0.405\,/\,0.401 & 0.271\,/\,0.251 & 0.557\,/\,0.587 & 0.569\,/\,0.519 \\
    SSIM &
    0.109\,/\,0.048 & 0.441\,/\,0.435 & 0.662\,/\,0.702 & 0.367\,/\,0.355 & 0.454\,/\,0.406 &
    0.412\,/\,0.373 & 0.206\,/\,0.190 & 0.636\,/\,0.636 & 0.579\,/\,0.536 \\
    VIF &
    0.152\,/\,0.125 & 0.379\,/\,0.360 & 0.506\,/\,0.564 & 0.151\,/\,0.133 & 0.326\,/\,0.239 &
    0.392\,/\,0.393 & 0.228\,/\,0.208 & 0.451\,/\,0.449 & 0.417\,/\,0.453 \\
    VIF$\mathrm{_{s0}}$ &
    0.105\,/\,0.109 & 0.367\,/\,0.348 & 0.484\,/\,0.539 & 0.118\,/\,0.102 & 0.307\,/\,0.216 &
    0.380\,/\,0.381 & 0.228\,/\,0.182 & 0.432\,/\,0.427 & 0.362\,/\,0.418 \\
    VIF$\mathrm{_{s1}}$ &
    0.234\,/\,0.269 & 0.466\,/\,0.398 & 0.776\,/\,0.788 & 0.458\,/\,0.381 & 0.547\,/\,0.473 &
    0.457\,/\,0.432 & 0.444\,/\,0.391 & 0.618\,/\,0.588 & 0.753\,/\,0.717 \\
    VIF$\mathrm{_{s2}}$ &
    0.298\,/\,0.301 & 0.489\,/\,0.414 & 0.807\,/\,0.811 & 0.546\,/\,0.462 & 0.611\,/\,0.533 &
    0.443\,/\,0.415 & 0.573\,/\,0.517 & 0.675\,/\,0.647 & 0.790\,/\,0.746 \\
    VIF$\mathrm{_{s3}}$ &
    \textbf{0.381\,/\,0.362} & 0.531\,/\,0.457 & 0.818\,/\,0.819 & 0.609\,/\,0.525 & 0.649\,/\,0.581 &
    0.439\,/\,0.407 & \textbf{0.640\,/\,0.606} & 0.716\,/\,0.701 & \textbf{0.807\,/\,0.756} \\
    ADM &
    0.212\,/\,0.176 & 0.628\,/\,0.709 & 0.858\,/\,0.906 & 0.758\,/\,0.782 & 0.690\,/\,0.659 &
    0.446\,/\,0.432 & 0.380\,/\,0.356 & 0.668\,/\,0.668 & 0.752\,/\,0.668 \\
    ADM$\mathrm{_{s0}}$ &
    0.184\,/\,0.092 & 0.304\,/\,0.352 & 0.569\,/\,0.616 & 0.155\,/\,0.199 & 0.238\,/\,0.170 &
    0.187\,/\,0.165 & 0.143\,/\,0.134 & 0.301\,/\,0.269 & 0.496\,/\,0.452 \\
    ADM$\mathrm{_{s1}}$ &
    0.161\,/\,0.161 & 0.458\,/\,0.501 & 0.782\,/\,0.823 & 0.432\,/\,0.462 & 0.538\,/\,0.488 &
    0.377\,/\,0.373 & 0.224\,/\,0.218 & 0.508\,/\,0.496 & 0.781\,/\,0.662 \\
    ADM$\mathrm{_{s2}}$ &
    0.211\,/\,0.183 & 0.646\,/\,0.705 & \textbf{0.883\,/\,0.918} & 0.772\,/\,0.807 & \textbf{0.717\,/\,0.684} &
    0.469\,/\,0.441 & 0.387\,/\,0.257 & 0.646\,/\,0.643 & 0.702\,/\,0.617 \\
    ADM$\mathrm{_{s3}}$ &
    0.277\,/\,0.196 & \textbf{0.757\,/\,0.723} & 0.841\,/\,0.869 & \textbf{0.877\,/\,0.901} & 0.692\,/\,0.668 &
    \textbf{0.482\,/\,0.450} & 0.465\,/\,0.368 & \textbf{0.706\,/\,0.731} & 0.682\,/\,0.605 \\
  \hline\hline
  \end{tabular}
  \end{table*}

\subsection{MOS Rate-Distortion Curves}
The Rate-Distortion (or Rate-Quality) curve is a common tool for comparing different encoders or encoding settings in lossy compression. In our study, we were able to collect the bitstream and subjective quality score for each distorted video. Given these results, we compared different \textit{chroma\_qp\_offset} ($\mathrm{\Delta QP_c}$) settings, as shown in Fig. \ref{fig:mos_rd}, by plotting MOS against bitrate. A key result in these RD-curves is that, when increasing $\mathrm{\Delta QP_c}$ by $6$ or $12$ at a high bitrate, roughly $2.5\%$--$17.2\%$ of bits can be reduced without suffering losses of perceptual quality. Occasionally, it may be observed that a video with small $\mathrm{\Delta QP_c}$ was slightly preferred over the setting $\mathrm{\Delta QP_c}=0$. This could be because both distorted videos were of very high quality, whereby some subjects were unable to distinguish differences in subjective quality, i.e., this may be attributed to statistical noise. On the other hand, the MOS sometimes dropped drastically even with slight increments of chroma quantization, at low bitrate settings. Towards understanding this, we observed more severe color shifts due to the additional quantization that affects the low frequency chroma components. It is possible that human perception is sensitive to this distortion type. Additionally, the percentage of bitrate reduction was not as significant as at high bitrates. This observation has important implication for encoding recipes: there is still room for perceptually optimizing coding efficiency, by better configuring VQA models to align with human percepts of chroma distortion.

\subsection{Limitations of the Current Study}
Introducing excessive chroma distortion is relatively new to the construction of a video quality dataset. Despite our best efforts, there are several limitations of this database. As with most the lab-controlled subjective studies, the experiment was hardly exhaustive. Although we designed a database having reasonably comprehensive content and distortion types, it is still circumscribed by several factors, such as the number of subjects and the quantity of subjective quality labels. As a result, constraints had to be made on the combinations of contents and distortions.

In our experiment design, we only included distortion types from luma and chroma compression. However, in real world applications, such as adaptive streaming, resolution changes are often used to optimize RD performance and can also cause distortions (scaling artifacts). Also, studying videos having distorted luma channels with pristine chroma channels may be of interest. Due to the lack of capacity, we did not include this distortion scenario in our dataset. Furthermore, recalling the preceding RD-curve examples in Fig. \ref{fig:mos_rd}, it was observed that bitrate reductions from heavier quantization of chroma only occurred in high bitrate regions. Yet, the granularity of luma distortion levels was limited, making it difficult to find a precise sweet spot to optimize a codec.

\section{Objective Video Quality Model Design}
VMAF is a data driven video quality prediction framework that extracts a number elementary VQA features then nonlinearly fuses these features by training a Support Vector Regressor (SVR). The current version of VMAF extracts the Additive Distortion Metric (ADM) \cite{LiADM2011} feature and four Visual Information Fidelity (VIF) \cite {SheikhB06} features computed on different oriented frequency bands. The Temporal Information (TI) feature of VMAF is simply the average difference between consecutive frames. This is used to capture temporal distortions associated with, or possibly causing motion or change. These six features are all derived on the luminance component only. Of course, this current limitation hinders VMAF from capturing chromatic distortions.

\subsection{Chromatic Features}
\begin{figure}[t]
  \centerline{
  \includegraphics[width=0.9\columnwidth]{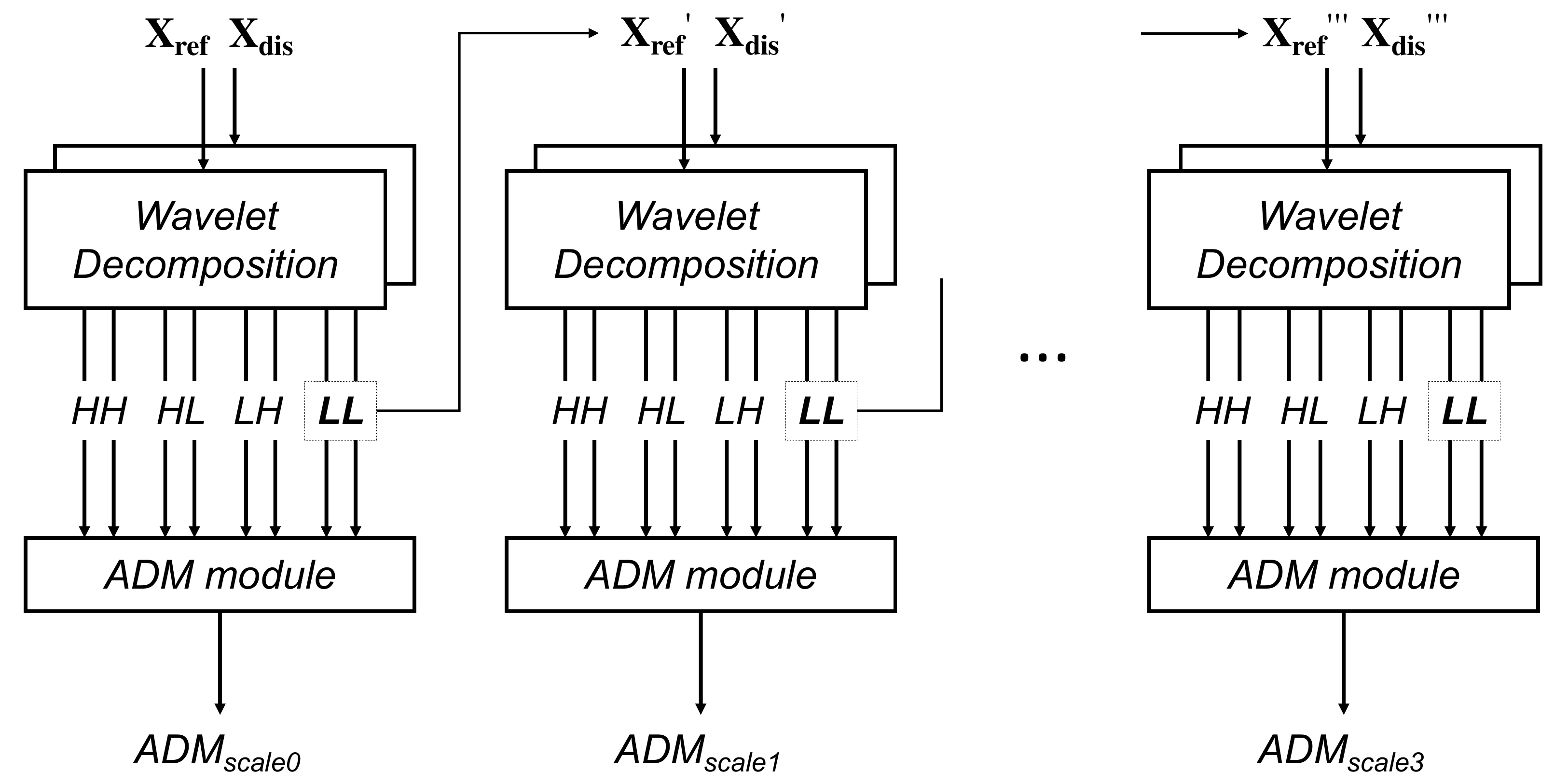}
  }
  \caption{Framework of multiscale ADM features. The ``LL" components from Wavelet transform in each scale are the input of the next, which produces different scales of ADM score iteratively. Larger scale index represents lower frequency subband.}
  \label{fig:adm_scale}
\end{figure}

Clearly, our goal is to find features that are expressive of chromatic distortions and that can be used to benefit the training of VQA models. To this end, we selected the VMAF features, along with the well-known PSNR and SSIM algorithms as candidates for the experiments on chroma distortion. We then conducted a systematic evaluation to understand the performances of these existing luma features and algorithms when applied on chroma channels. It should be noted that the chroma channels only have halved horizontal and vertical resulutions compared to luma, since we used videos with YUV420 format for all the experiments. Table \ref{tab:chroma_feature_srcc} tabulates the obtained Spearman Rank Order Correlation Coefficients (SROCC) obtained on nine databases, including NFLX$_\text{c}$. The first thing to notice is that the two most widely-used perceptual quality models (SSIM and VIF) performed far below desired levels. This was possibly due to the fact that they have been designed and extensively tested only on luminance signals. However, it must also be recognized that chromatic components possess different statistical and scaling properties than luma components. Interestingly, the third scale of the ADM features (ADM$^\text{(Cb)}_\text{s3}$ and ADM$^\text{(Cr)}_\text{s3}$) achieved standout performance across most of the databases. In particular, on NFLX$_\text{c}$, it achieved close to $0.9$ of SROCC. This suggests that this feature is highly consistent with the human perception context of chromatic distortion. We also illustrate the multiscale framework of the ADM feature in Fig. \ref{fig:adm_scale} for better understanding.

The second observation that can be made is with regards to the multiscale behavior of VIF and ADM: performance was improved as the scale index was increased (lower frequency subband). This is not surprising, since the human chromatic contrast sensitivity functions (CSF) (red-green and blue-yellow color opponents) \cite{Mullen85, Sekiguchi1993, Rovamo1999} pass much lower ranges of frequencies than the luminance CSF. The human visual system neglects higher frequencies when processing chromatic information. 

\subsection{Learning a Color VMAF model}
We began by empirically including ADM$^\text{(Cb)}_\text{s3}$ and ADM$^\text{(Cr)}_\text{s3}$ as two additional features for training chroma-aware VMAF models, given its exceptional performance on NFLX$_\text{c}$. However, while the trained model significantly boosted correlation on NFLX$_\text{c}$, we discerned considerable performance degradation on other databases, especially LIVE-VQA. In fact, when tested on LIVE-VQA, all of the chromatic features yielded unsatisfactory SROCC, as may be seen in Table \ref{tab:chroma_feature_srcc}. This may have been because nearly half of the LIVE-VQA database videos contain transient distortions produced by simulated network losses, which may not be effectively captured by chromatic features. It is important to note that these distinct distortions are different from our training data (see Section \ref{sec:criteria}), posing additional challenges. In any case, without careful training, the additional chroma features can lower performance.

\begin{table}[t]
  \caption{SROCC performance with respect to different quantization step sizes represented by $N$. The last row ($N=\infty$) denotes the features without quantization. Please refer to the dataset acronyms in section \ref{sec:dataset}.}
  \centering
  \scalebox{1.0}{
    \renewcommand{\tabcolsep}{5pt} 
    \renewcommand{\arraystretch}{1.07} 
  \begin{tabular}{cccccc}
  \hline \hline
  N        & L-VQA & NFLX$_\text{c}$ & EPFL & VQEG & Overall \\\hline
  4        & 0.744 & 0.772 & 0.879 & 0.829 & 0.829\\
  8        & 0.740 & 0.822 & 0.873 & 0.834 & \textbf{0.838}\\
  16       & 0.706 & 0.862 & 0.858 & 0.829 & 0.836\\
  32       & 0.701 & 0.875 & 0.845 & 0.830 & 0.831\\
  64       & 0.703 & 0.877 & 0.841 & 0.828 & 0.831\\
  $\infty$ & 0.692 & 0.880 & 0.841 & 0.827 & 0.830\\ \hline \hline
  \end{tabular}}
  \label{feature_q_step}
\end{table}

In order to tackle this issue, we regularized the additional chromatic features using a uniform quantization function. A uniform quantizer with a parameterized quantization step size $\Delta_N = 1/N$ that maps a real value $x\in(0,1]$ to $N$ discrete values is given by
\begin{equation}\label{quantization_eq}
  \begin{split}
    \tilde{x} = Q_N(x)=\Delta_N \lceil \frac{x}{\Delta_N}\rceil.
  \end{split}
\end{equation}
The spirit behind this approach is simple: when a feature is excessively quantized, say, with $N=1$, the feature becomes a constant to the regression model. In this case, the quantized variable cannot contribute to the model learning, resulting in the same feature set as VMAF 0.6.1; conversely, as $N$ grows, the result becomes closer to the original feature. Thus, the quantization function allows a flexible trade off in performance on NFLX$_\text{c}$ and the other databases. We studied the effect of using different step sizes $\Delta_N$, and report the results in Table \ref{feature_q_step}. It may be seen that the SROCC performance on LIVE-VQA and EPFL decreased as $N$ was increased, whereas a reversed trend was observed on NFLX$_\text{c}$. Some databases, such as VQEG, were less affected by varying step size. Based on the results of Table \ref{feature_q_step}, we chose $N=8$ to quantize the chroma ADM features.

Ultimately, the feature vector used to learn the color VMAF model, which we refer to as VMAF$_\text{c}$, is expressed by
\begin{equation}
  \begin{split}
    \mathbf{f}_{\text{VMAF}_\text{c}} = [&\text{VIF}_\text{s0}, \text{VIF}_\text{s1}, \text{VIF}_\text{s2}, \text{VIF}_\text{s3},
    \text{TI}, \\ &\text{ADM},\tilde{\text{ADM}}^\text{(Cb)}_\text{s3},\tilde{\text{ADM}}^\text{(Cr)}_\text{s3}].
  \end{split}
\end{equation}
The terms $\tilde{\text{ADM}}^\text{(Cb)}_\text{s3}$ and $\tilde{\text{ADM}}^\text{(Cr)}_\text{s3}$ denote the ${\text{ADM}}^\text{(Cb)}_\text{s3}$ and ${\text{ADM}}^\text{(Cr)}_\text{s3}$ features quantized by (\ref{quantization_eq}). The finalized VMAF$_\text{c}$ model was trained on the VMAF+ dataset \cite{bampis2018spatiotemporal}, as we will describe in section \ref{subs:overall}.

\begin{table*}[!t]
  \renewcommand{\arraystretch}{1.3}
  \caption{Overall performance comparison of VQA algorithms: Each cell shows the SROCC and PLCC values of an VQA model evaluated on a specific database, expressed as SROCC\,/\,PLCC. Both VMAF 0.6.1 and VMAF$_\text{c}$ were trained on the VMAF+ dataset \cite{bampis2018spatiotemporal}. The three best SROCC results among each dataset are denoted by boldface.}
    \label{tab:overall_comparison}
    \centering
    \renewcommand{\tabcolsep}{2.7pt} 
    \begin{tabular}{l c c c c c c c c c c}
    \hline\hline
    Dataset & 
    LIVE-VQA & LIVE-MBL & NFLX & \textbf{NFLX}$\mathbf{_c}$ & BVI-HD & CSIQ-VQA & EPFL & VQEG & SHVC & Overall \\
    \hline
    PSNR$_\text{Y}$ &
    0.523\,/\,0.549 & 0.687\,/\,0.717 & 0.705\,/\,0.706 & 0.412\,/\,0.424 & 0.588\,/\,0.600 & 0.579\,/\,0.565 & 0.753\,/\,0.754 & 0.770\,/\,0.776 & 0.755\,/\,0.761 & 0.655\,/\,0.664 \\
    PSNR$_\text{411}$ &
    0.434\,/\,0.465 & 0.663\,/\,0.690 & 0.703\,/\,0.698 & 0.571\,/\,0.585 & 0.565\,/\,0.577 & 0.545\,/\,0.536 & 0.598\,/\,0.613 & 0.734\,/\,0.728 & 0.738\,/\,0.746 & 0.626\,/\,0.635 \\
    PSNR$_\text{611}$ &
    0.459\,/\,0.494 & 0.672\,/\,0.698 & 0.704\,/\,0.702 & 0.529\,/\,0.546 & 0.573\,/\,0.585 & 0.555\,/\,0.544 & 0.644\,/\,0.653 & 0.745\,/\,0.743 & 0.749\,/\,0.755 & 0.635\,/\,0.644 \\
    CSPSNR &
    0.498\,/\,0.530 & 0.685\,/\,0.714 & 0.704\,/\,0.705 & 0.588\,/\,0.605 & 0.581\,/\,0.592 & 0.572\,/\,0.558 & 0.727\,/\,0.724 & 0.765\,/\,0.767 & 0.750\,/\,0.757 & 0.661\,/\,0.670 \\
    PSNR$_\text{HVS}$ &
    0.662\,/\,0.689 & 0.757\,/\,0.784 & 0.819\,/\,0.810 & 0.516\,/\,0.534 & 0.739\,/\,0.748 & 0.599\,/\,0.641 & 0.904\,/\,0.909 & 0.798\,/\,0.799 & 0.831\,/\,0.872 & 0.758\,/\,0.776 \\
    SSIM &
    0.694\,/\,0.704 & 0.757\,/\,0.767 & 0.788\,/\,0.790 & 0.555\,/\,0.560 & \textbf{0.784}\,/\,0.786 & 0.698\,/\,0.712 & 0.712\,/\,0.703 & 0.907\,/\,0.909 & 0.754\,/\,0.817 & 0.754\,/\,0.766 \\
    MS-SSIM &
    0.732\,/\,0.739 & 0.748\,/\,0.761 & 0.741\,/\,0.745 & 0.524\,/\,0.525 & 0.747\,/\,0.752 & \textbf{0.749}\,/\,0.746 & \textbf{0.931}\,/\,0.934 & 0.898\,/\,0.902 & 0.715\,/\,0.787 & 0.780\,/\,0.791 \\
    ST-RRED &
    \textbf{0.802}\,/\,0.801 & 0.881\,/\,0.903 & 0.762\,/\,0.761 & 0.616\,/\,0.612 & \textbf{0.781}\,/\,0.797 & \textbf{0.801}\,/\,0.786 & \textbf{0.950}\,/\,0.952 & \textbf{0.921}\,/\,0.708 & \textbf{0.888}\,/\,0.899 & \textbf{0.846}\,/\,0.829 \\
    SpEED-QA &
    0.767\,/\,0.763 & \textbf{0.883}\,/\,0.905 & 0.780\,/\,0.781 & \textbf{0.622}\,/\,0.626 & 0.770\,/\,0.784 & 0.746\,/\,0.747 & \textbf{0.941}\,/\,0.869 & \textbf{0.908}\,/\,0.659 & 0.880\,/\,0.887 & 0.834\,/\,0.798 \\
    ST-MAD &
    \textbf{0.825}\,/\,0.830 & 0.663\,/\,0.686 & 0.768\,/\,0.746 & 0.549\,/\,0.550 & 0.757\,/\,0.758 & 0.735\,/\,0.740 & 0.901\,/\,0.908 & 0.847\,/\,0.840 & 0.611\,/\,0.619 & 0.760\,/\,0.761 \\
    VQM-VFD &
    \textbf{0.804}\,/\,0.823 & 0.816\,/\,0.847 & \textbf{0.931}\,/\,0.942 & 0.597\,/\,0.624 & \textbf{0.792}\,/\,0.802 & \textbf{0.839}\,/\,0.830 & 0.850\,/\,0.847 & \textbf{0.939}\,/\,0.943 & 0.863\,/\,0.888 & \textbf{0.847}\,/\,0.859 \\
    iCID &
    0.552\,/\,0.569 & 0.763\,/\,0.773 & 0.776\,/\,0.769 & \textbf{0.890}\,/\,0.892 & 0.709\,/\,0.714 & 0.679\,/\,0.682 & 0.778\,/\,0.774 & 0.887\,/\,0.893 & 0.717\,/\,0.728 & 0.766\,/\,0.773 \\
    VMAF$_\text{0.6.1}$ &
    0.752\,/\,0.759 & \textbf{0.905}\,/\,0.924 & \textbf{0.931}\,/\,0.944 & 0.612\,/\,0.627 & 0.772\,/\,0.785 & 0.615\,/\,0.624 & 0.844\,/\,0.858 & 0.857\,/\,0.866 & \textbf{0.901}\,/\,0.922 & 0.826\,/\,0.844 \\
    VMAF$_\text{c}$ &
    0.740\,/\,0.751 & \textbf{0.881}\,/\,0.901 & \textbf{0.932}\,/\,0.949 & \textbf{0.821}\,/\,0.832 & 0.759\,/\,0.766 & 0.597\,/\,0.623 & 0.873\,/\,0.883 & 0.834\,/\,0.852 & \textbf{0.912}\,/\,0.925 & \textbf{0.838}\,/\,0.855 \\
  \hline\hline
  \end{tabular}
  \end{table*}

\begin{figure*}[!t]
  \centering
  \footnotesize
  \renewcommand{\tabcolsep}{1.5pt} 
  \renewcommand{\arraystretch}{1} 
  \def\imgwid{0.325\textwidth}  
  \begin{tabular}{ccc}
  \includegraphics[width=\imgwid]{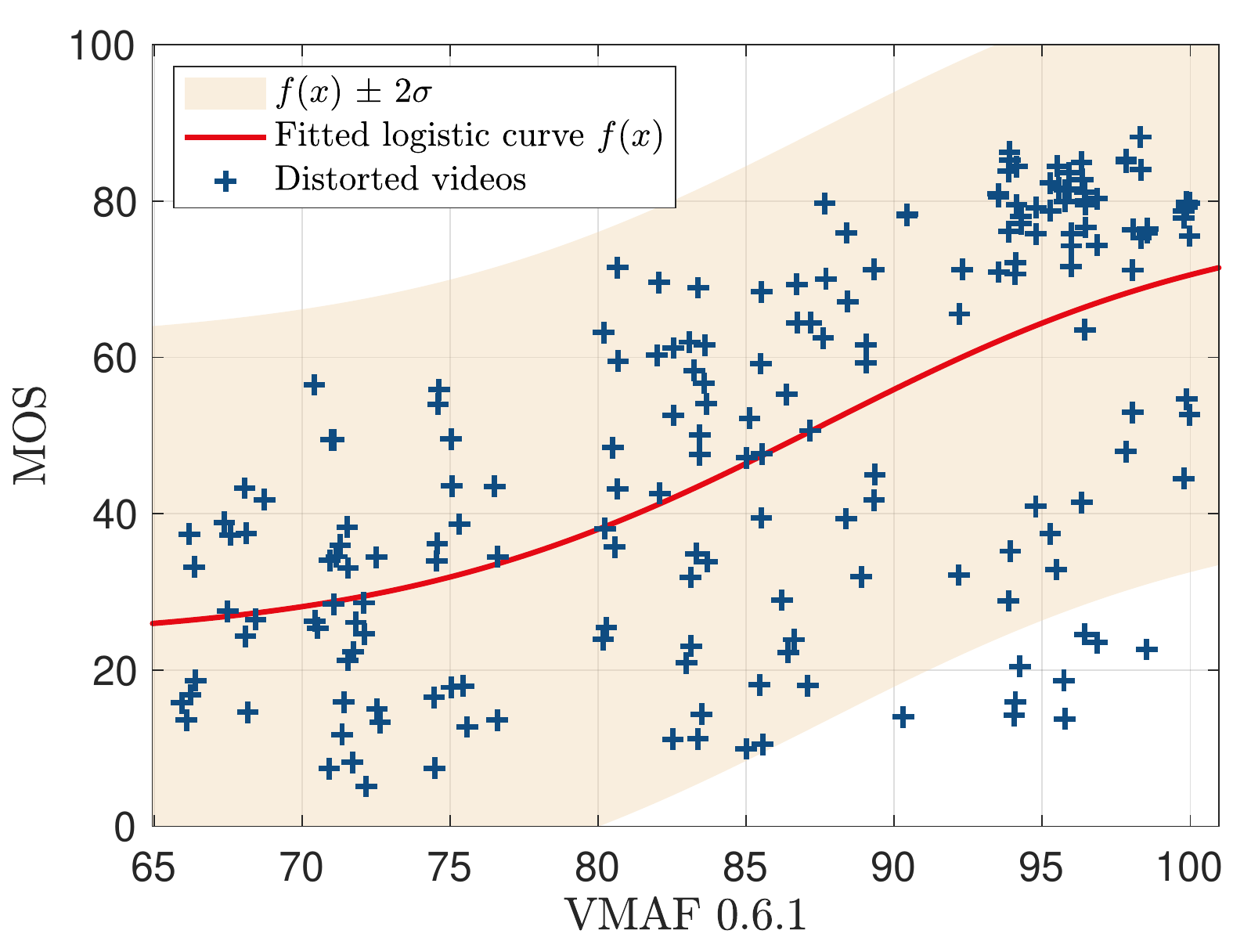} &
  \includegraphics[width=\imgwid]{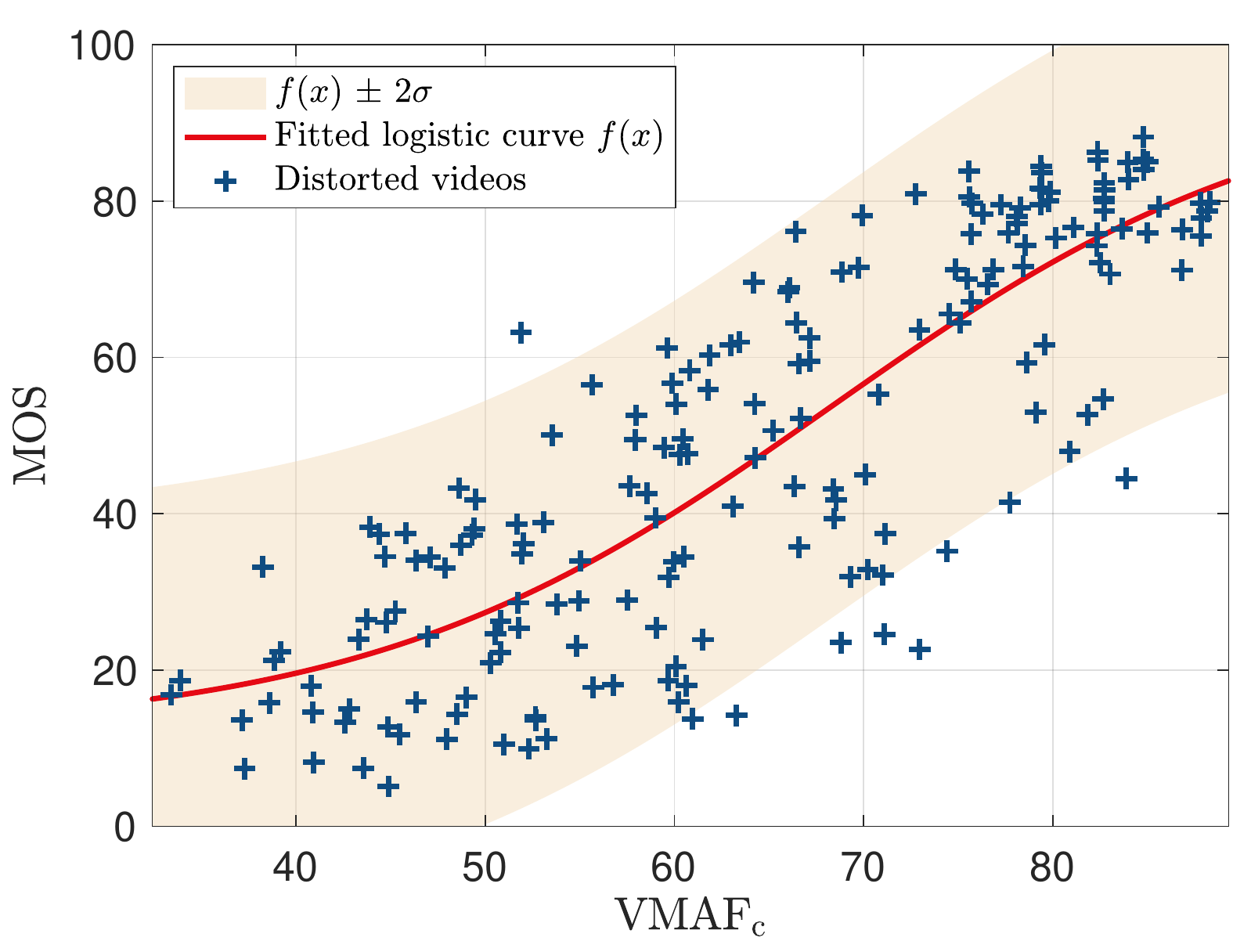} &
  \includegraphics[width=\imgwid]{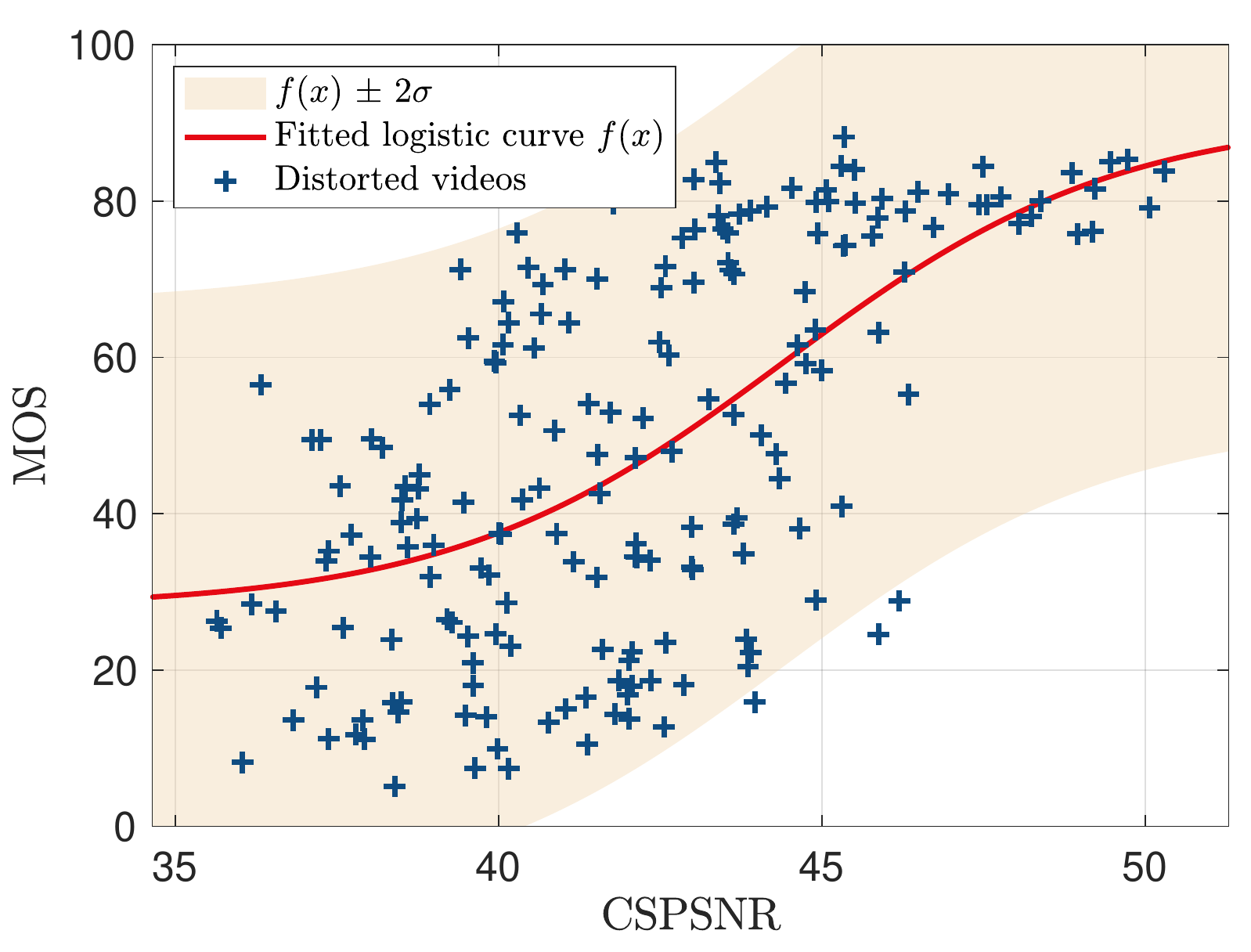} \\
  \end{tabular}
  \caption{Scatter plots and logistic regression curves of VMAF 0.6.1\,/\,VMAF$_\text{c}$\,/\,CSPSNR versus MOS on NFLX$_\text{c}$ dataset. The $\sigma$ denotes the standard deviation of the residuals between MOS values and $f(x)$. We follow the style in \cite{tu2020bband}}.
  \label{fig:scatter_plots}
\end{figure*}

\section{Objective Video Quality Assessment Experiments}
\subsection{Experimental Setup}\label{sec:criteria}
\subsubsection{Evaluation Datasets}\label{sec:dataset}
In addition to NFLX$_\text{c}$, we tested the models on a variety of subjective VQA databases, including LIVE VQA \cite{Seshadrinathan2010}, LIVE Mobile (LIVE-MBL)\footnote[4]{We only used the mobile subset in LIVE Mobile} \cite{Moorthy2012}, NFLX \cite{ZliVMAF16}, CSIQ VQA \cite{Vu2014}, BVI-HD \cite{ZhangBVIHD2018}, VQEG-HD3 \cite{VQEG_DB}, EPFLPolimi \cite{DeSimoneEPFL2010}, and SHVC \cite{SHVC_DB}. These publicly available datasets are commonly used to evaluate VQA models. They contain a large variety of contents, resolutions, and distortion types, such as MPEG4/H.264/HEVC compression, resolution changes, transmission errors, frame rate adaptations, and so on.
\subsubsection{Evaluation Criteria}
To evaluate the performance of a VQA model, we used the SROCC and the Pearson linear correlation coefficient (PLCC), which are calculated between the ground truth MOS and the predicted scores. The SROCC measures the degree of monotonic relationship between two variables, while the PLCC is computed after a logistic mapping \cite{logistic_vqeg} to measure the degree of linear correlation against MOS. Larger values of SROCC\,/\,PLCC indicate better performance in terms of correlation with human perception. To compute the overall correlation, we applied Fisher’s $z$-transform \cite{itup1401} to each correlation coefficient value $r$
\begin{equation}\label{eq:fisher}
  z = \frac{1}{2}\ln \frac{1+r}{1-r},
\end{equation}
then averaged them over all the test databases \cite{bampis2018spatiotemporal}. This average value of transformed correlation coefficients is then transformed back using the inverse function $r = \tanh \left(z\right)$.
\subsubsection{Implementation Details}
We used the \textit{libsvm} package \cite{Chang2011libsvm} to implement $\nu$-SVR \cite{Schlkopf2000}. The effectiveness of SVR depends on the selection of kernel, for which we chose the radial basis function (RBF) with penalty parameter $C = 2^3$ and kernel parameter $\gamma = 2^{-3}$. The parameters were empirically selected from a 2D grid of values $(C,\gamma) \in \{2^{-5}, 2^{-4},...,2^5\}\times\{2^{-5}, 2^{-4},...,2^5\}$. Instead of maximizing the overall performance, we selected a sub-optimal parameter set satisfying a monotonicity property (see Section \ref{sec:mono}), as well as ensuring an appropriate trade-off between NFLX$_\text{c}$ and the other datasets. Before training or inferencing, all of the features were linearly rescaled to $[0,1]$ using min-max normalization. For the sake of simplicity and stability, we used the arithmetic mean \cite{tu2020pooling} to aggregate the per-frame scores.

\begin{table*}[tp]
  \renewcommand{\tabcolsep}{2pt} 
  \renewcommand{\arraystretch}{1.2} 
  \caption{Results of significance tests between the performances of VQA models on the NFLX$_\text{c}$ dataset. Each cell shows the statistical significance of SROCC and PLCC. A value of `1' indicates that the row has statistically higher correlation coefficient than the column, while `0' signifies that the column has statistically lower correlation coefficient than the row. A symbol of `-' indicates no statistical difference between the correlation coefficients of row and column.}
  \centering
  \scriptsize
  \begin{tabular}{l c c c c c c c c c c c c c c}
  \hline\hline
        & PSNR$_\text{Y}$ & PSNR$_\text{411}$ & PSNR$_\text{611}$ & CSPSNR & PSNR$_\text{HVS}$ & SSIM & MS-SSIM & ST-RRED & SpEED-QA & ST-MAD & VQM-VFD & iCID & VMAF$_\text{0.6.1}$ & VMAF$_\text{c}$\\ \hline
  PSNR$_\text{Y}$ & ~~~~-\,-~~~~ & ~~~~0\,0~~~~ & ~~~~-\,-~~~~ & ~~~~0\,0~~~~ & ~~~~-\,-~~~~ & ~~~~-\,-~~~~ & ~~~~-\,-~~~~ & ~~~~0\,0~~~~ & ~~~~0\,0~~~~ & ~~~~-\,-~~~~ & 0\,0 & ~~~~0\,0~~~~ & ~~~~0\,0~~~~ & ~~~~0\,0~~~~ \\
  PSNR$_\text{411}$ & 1\,1 & -\,- & -\,- & -\,- & -\,- & -\,- & -\,- & -\,- & -\,- & -\,- & -\,- & 0\,0 & -\,- & 0\,0\\
  PSNR$_\text{611}$ & -\,- & -\,- & -\,- & -\,- & -\,- & -\,- & -\,- & -\,- & -\,- & -\,- & -\,- & 0\,0 & -\,- & 0\,0\\
  CSPSNR & 1\,1 & -\,- & -\,- & -\,- & -\,- & -\,- & -\,- & -\,- & -\,- & -\,- & -\,- & 0\,0 & -\,- & 0\,0\\
  PSNR$_\text{HVS}$ & -\,- & -\,- & -\,- & -\,- & -\,- & -\,- & -\,- & -\,- & -\,- & -\,- & -\,- & 0\,0 & -\,- & 0\,0\\
  SSIM & -\,- & -\,- & -\,- & -\,- & -\,- & -\,- & -\,- & -\,- & -\,- & -\,- & -\,- & 0\,0 & -\,- & 0\,0\\
  MS-SSIM & -\,- & -\,- & -\,- & -\,- & -\,- & -\,- & -\,- & -\,- & -\,- & -\,- & -\,- & 0\,0 & -\,- & 0\,0\\
  ST-RRED & 1\,1 & -\,- & -\,- & -\,- & -\,- & -\,- & -\,- & -\,- & -\,- & -\,- & -\,- & 0\,0 & -\,- & 0\,0\\
  SpEED-QA & 1\,1 & -\,- & -\,- & -\,- & -\,- & -\,- & -\,- & -\,- & -\,- & -\,- & -\,- & 0\,0 & -\,- & 0\,0\\
  ST-MAD & -\,- & -\,- & -\,- & -\,- & -\,- & -\,- & -\,- & -\,- & -\,- & -\,- & -\,- & 0\,0 & -\,- & 0\,0\\
  VQM-VFD & 1\,1 & -\,- & -\,- & -\,- & -\,- & -\,- & -\,- & -\,- & -\,- & -\,- & -\,- & 0\,0 & -\,- & 0\,0\\
  iCID & 1\,1 & 1\,1 & 1\,1 & 1\,1 & 1\,1 & 1\,1 & 1\,1 & 1\,1 & 1\,1 & 1\,1 & 1\,1 & -\,- & 1\,1 & 1\,1\\
  VMAF$_\text{0.6.1}$ & 1\,1 & -\,- & -\,- & -\,- & -\,- & -\,- & -\,- & -\,- & -\,- & -\,- & -\,- & 0\,0 & -\,- & 0\,0\\
  VMAF$_\text{c}$ & 1\,1 & 1\,1 & 1\,1 & 1\,1 & 1\,1 & 1\,1 & 1\,1 & 1\,1 & 1\,1 & 1\,1 & 1\,1 & 0\,0 & 1\,1 & -\,-\\
  \hline\hline
  \end{tabular}
  \label{tab:significance_test}
\end{table*}

\subsection{Overall Comparison}\label{subs:overall}
We evaluated the optimized VMAF$_\text{c}$ against a number of
popular FR IQA/VQA models: PSNR, PSNR$_\text{411}$, PSNR$_\text{611}$, CSPSNR \cite{Shang2019}, PSNR-HVS-M \cite{psnrhvsm07}, SSIM \cite{WangBSS04}, MS-SSIM \cite{WangMSSSIM03}, ST-RRED \cite{Soundararajan2013}, SpEED-QA \cite{BampisSpeed2017}, ST-MAD \cite{Vu2011}, VQM-VFD \cite{Pinson2014}, iCID \cite{Preiss2014}, and VMAF 0.6.1 \cite{ZliVMAF16}. Since separate databases cannot be combined into one in a simple way, most learning-based IQA/VQA models are train/test on multiple databases independently with cross validation. Under this setting, each dataset is typically split into 80\%-20\% portions with respect to content for training-testing, and the model parameters are searched to maximize individual correlations. Nonetheless, this does not give a general model that can be practically used and is prone to biases from database-specific characteristics. To avoid these problems, we stayed in line with the benchmark criteria of VMAF, where the models were trained on a \textit{distinct} dataset VMAF+ \cite{bampis2018spatiotemporal} and evaluated on the others. VMAF+ is a large scale VQA dataset designed for training VMAF, containing 522 distorted videos subjected to different levels of scaling and compression artifacts. The performance results are shown in Table \ref{tab:overall_comparison} and plots of model predictions versus MOS are shown in Fig. \ref{fig:scatter_plots}. With further investigation, we observed that the outliers in the scatter plot of VMAF 0.6.1 are mostly the videos with $\left(\text{QP}_\text{Y},\text{QP}_\text{c}\right)=\left(15,51\right)$, where chroma distortions were decoupled from luma. These are the instances which VMAF 0.6.1 failed the most.

From these results, we can draw a number of interesting conclusions. The first thing to notice is the performance on the NFLX$_\text{c}$ dataset. The best performer among all the VQA models, except VMAF$_\text{c}$ and iCID, only reaches SROCC slightly higher than $0.6$, whereas VMAF$_\text{c}$ outperformed most of the other methods, achieving $0.82$ SROCC. This improvement over VMAF 0.6.1 is understandable, since the chromatic features were efficiently integrated. It may also be found that VMAF$_\text{c}$ performed marginally worse than VMAF 0.6.1 on some databases. Overall, a gain of about 0.01 in both SROCC\,/\,PLCC was achieved by VMAF$_\text{c}$. Regarding the training set of VMAF$_\text{c}$, it should be noted that the VMAF+ dataset does not incorporate any videos with the setting of $\Delta\text{QP}_{c}$, yet a remarkable performance improvement is still attained on NFLX$_\text{c}$. This is quite significant, since unlike other databases, NFLX$_\text{c}$ allows the measurement of performance on compressed videos with independent chroma compression, which as we have shown, can result in significantly improved perceptual rate-distortion optimization. Despite being the top-performer on NFLX$_\text{c}$, the iCID model failed on many of the other datasets with an overall SROCC performance of $0.766$, making it hard to justify its use in practical applications.

When comparing PSNR$_\text{Y}$, PSNR$_\text{411}$, PSNR$_\text{611}$, and CSPSNR, it may be observed that the levels of performance attained by the PSNR family are quite poor. However, when tested on NFLX$_\text{c}$, CSPSNR did better than PSNR$_\text{411}$ and PSNR$_\text{611}$, which promotes the result reported in \cite{Shang2019}.

\subsection{Significance Test}
We further analyzed the statistical significance of model performances as expressed by SROCC and PLCC reported in Table \ref{tab:overall_comparison}, following the recommended procedure in section 7.6.1 of ITU-T Rec. P.1401 \cite{itup1401}. The test uses statistics derived from Fisher’s-$z$ transformed correlation coefficients in each comparison, compared with the $95\%$ two-tailed Student's t-test critical value. Table \ref{tab:significance_test} shows the results of the statistical significance tests.

From the results shown in the table, we may observe that iCID and VMAF$_\text{c}$ statistically surpassed all the other models, since most of them only utilize luminance information. Unsurprisingly, PSNR$_\text{Y}$ performed significantly worse than most of the models.
It may also be noticed that there was no statistical difference observed when comparing the other models. This is likely because the test methodology is too conservative, given the limited sample size used to calculate correlation coefficients. However, the VMAF$_\text{c}$ model was still statistically better than most of the other models under this protocol.

\begin{table}[t]
  \caption{Comparison against deep learning based VQA model.}
  \centering
  \scalebox{1.0}{
    \renewcommand{\tabcolsep}{4.5pt} 
    \renewcommand{\arraystretch}{1.2} 
  \begin{tabular}{l cc c cc }
  \hline\hline    
                  & \multicolumn{2}{c}{LIVE-VQA} & & \multicolumn{2}{c}{CSIQ-VQA} \\ 
                  \cline{2-3}\cline{5-6}
                  & SROCC & PLCC~ & & SROCC & PLCC~ \\ \hline
  DeepVQA         & 0.891 & 0.881 & & 0.904 & 0.901 \\
  VMAF            & 0.939 & 0.915 & & 0.635 & 0.575 \\
  VMAF$_\text{c}$ & 0.934 & 0.918 & & 0.683 & 0.632 \\ \hline\hline
  \end{tabular}}
  \label{tab:compare_deep}
\end{table}

\begin{table*}[!t]
  \renewcommand{\arraystretch}{1.3}
  \caption{Cross dataset comparison of the VMAF$_\text{c}$ model. Each cell shows the SROCC performance of training on the dataset in the row and testing on the dataset in the column. The best overall performance is highlighted in boldface.}
    \label{tab:cross_comparison}
    \centering
    \renewcommand{\tabcolsep}{1.0pt} 
    \begin{tabular}{l c c c c c c c c c c c}
    \hline\hline
    Dataset & 
    LIVE-VQA & LIVE-MBL & VMAF+ & NFLX & NFLX$_\text{c}$ & BVI-HD & CSIQ-VQA & EPFL & VQEG & SHVC & Overall \\
    \hline
    LIVE-VQA\phantom{0} &
    --    & 0.814 & 0.796 & 0.877 & 0.572 & 
    0.775 & 0.644 & 0.828 & 0.737 & 0.830 & 0.778 \\
    LIVE-MBL &
    0.654 & --    & 0.849 & 0.915 & 0.859 & 
    0.763 & 0.627 & 0.853 & 0.786 & 0.889 & 0.832 \\
    VMAF+ &
    0.706 & 0.891 & --    & 0.932 & 0.871 & 
    0.757 & 0.626 & 0.840 & 0.828 & 0.923 & \textbf{0.854} \\
    NFLX &
    0.723 & 0.916 & 0.903 & --    & 0.814 &
    0.784 & 0.575 & 0.870 & 0.844 & 0.886 & 0.851 \\
    NFLX$_\text{c}$ & 
    0.643 & 0.845 & 0.899 & 0.886 & --    &
    0.715 & 0.649 & 0.749 & 0.784 & 0.827 & 0.818 \\
    BVI-HD &
    0.648 & 0.841 & 0.821 & 0.906 & 0.797 & 
    --    & 0.642 & 0.876 & 0.797 & 0.863 & 0.813 \\
    CSIQ-VQA &
    0.527 & 0.832 & 0.823 & 0.799 & 0.874 &
    0.738 & --    & 0.775 & 0.799 & 0.829 & 0.785 \\
    EPFL &
    0.706 & 0.763 & 0.647 & 0.854 & 0.629 & 
    0.769 & 0.636 & --    & 0.726 & 0.860 & 0.766 \\
    VQEG &
    0.694 & 0.886 & 0.880 & 0.933 & 0.772 & 
    0.779 & 0.621 & 0.854 & --    & \phantom{00.}0.907\phantom{00.} & 0.832 \\
    SHVC &
    \phantom{00.}0.608\phantom{00.} & \phantom{00.}0.797\phantom{00.} & \phantom{00.}0.767\phantom{00.} & \phantom{00.}0.886\phantom{00.}  & \phantom{00.}0.857\phantom{00.} &
    \phantom{00.}0.749\phantom{00.} & \phantom{00.}0.632\phantom{00.} & \phantom{00.}0.686\phantom{00.} & \phantom{00.}0.696\phantom{00.} & --    & \phantom{.00}0.772\phantom{00.} \\
  \hline\hline
  \end{tabular}
  \end{table*}

\subsection{Comparison with a Deep Learning Based VQA Model}
Recently, deep convolutional neural networks have been shown to deliver standout performance on a wide variety of applications. In the field of video quality, a full-reference model called DeepVQA \cite{kim2018eccv} has been proposed, that achieves state-of-the-art performance on the LIVE-VQA and CSIQ-VQA datasets. Unfortunately, the authors could not provide a trained model that can be tested on all the VQA datasets. To fairly compare DeepVQA against VMAF\,/\,VMAF$_\text{c}$, we followed the train-test split reported in \cite{kim2018eccv}, and re-trained the VMAF models on each dataset. We used the same parameters $(C,\gamma)$ as the original model for simplicity. The results of the performance comparison are shown in Table \ref{tab:compare_deep}. Overall, DeepVQA yielded slightly worse SROCC\,/\,PLCC performance than VMAF and VMAF$_\text{c}$ on LIVE-VQA, while performing much better on CSIQ-VQA. This is because CSIQ-VQA mostly contains legacy distortions such as additive white noise (AWGN) or simulated wireless transmission loss, which are not of interest in modern video streaming scenarios. The results also indicate that deep neural networks have the potential to learn good features, including chroma, for assessing video quality.

\subsection{Cross-database Comparison}
In addition to analyzing model performance on one training dataset, we investigated the effects of using different datasets to train VMAF$_\text{c}$, with results reported in Table \ref{tab:cross_comparison}. Using the $10$ available VQA databases, we trained the SVR model (with feature $\mathbf{f}_{\text{VMAF}_\text{c}}$) on each dataset, then tested on the others. Due to the differences between the datasets, using the same parameters $(C,\gamma)=(2^3,2^{-3})$ yielded unsatisfactory performance on some training sets. Therefore, for fair comparison, we separately searched the SVR parameters on a $7\times7$ grid on each dataset to optimize the overall performance. The experimental results clearly shows the outstanding performance attained by using VMAF+ as the training data. Also, the SROCC attained when testing on NFLX$_\text{c}$ was generally greater than $0.7$, among the different training sets. This strongly suggests the robustness of the added chromatic features. It should be noted that the performance results on the VMAF+ training set differ slightly from the results reported in Table \ref{tab:overall_comparison}, due to different optimization objectives.

\begin{figure}[!t]
  \centering
  \footnotesize
  \renewcommand{\tabcolsep}{1.5pt} 
  \renewcommand{\arraystretch}{1} 
  \def\imgwid{0.238\textwidth}
  
  \begin{tabular}{cc}
    \includegraphics[width=\imgwid]{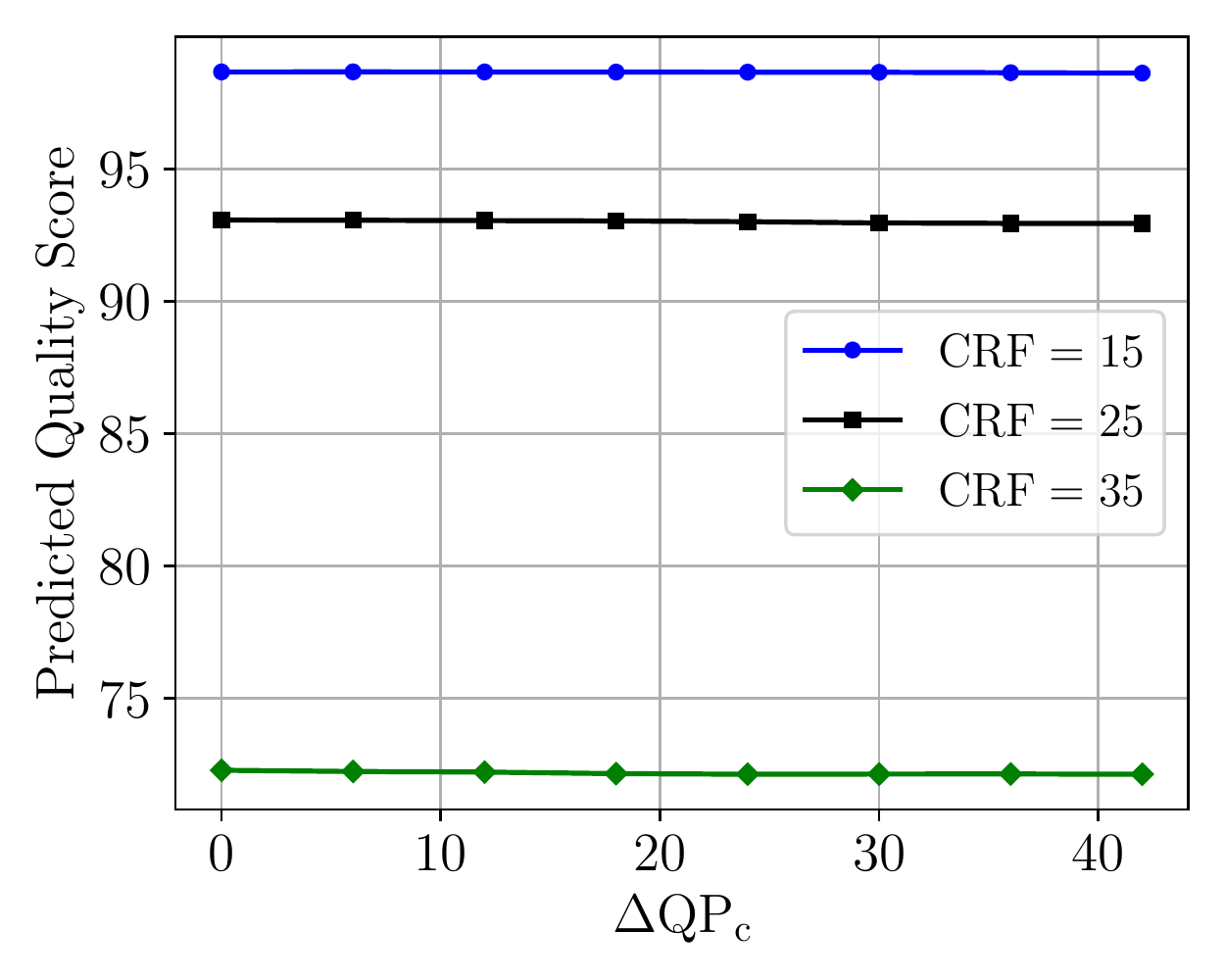} &
    \includegraphics[width=\imgwid]{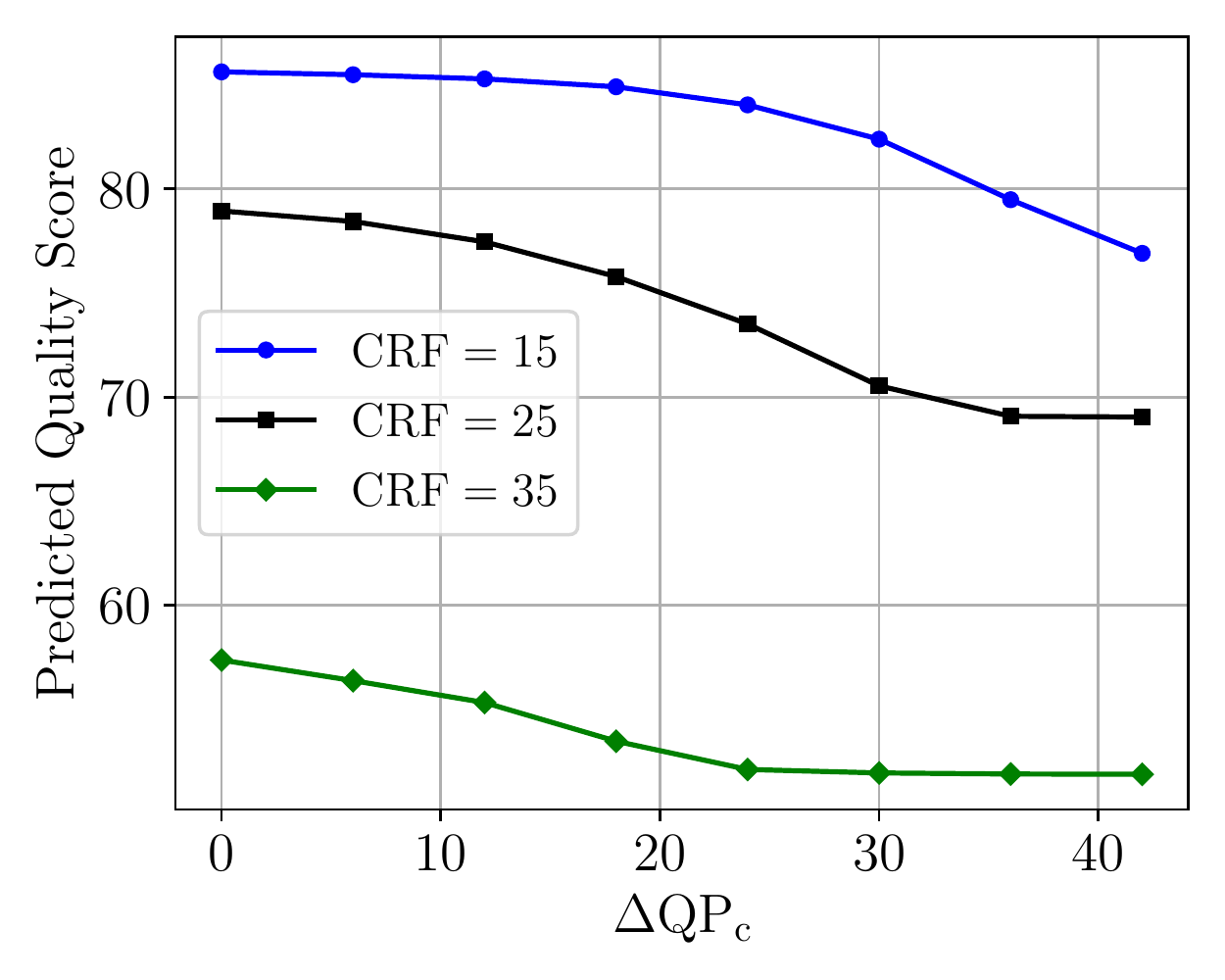} \\
    (a) VMAF 0.6.1 & (b) VMAF$_\text{c}$ \\
    \includegraphics[width=\imgwid]{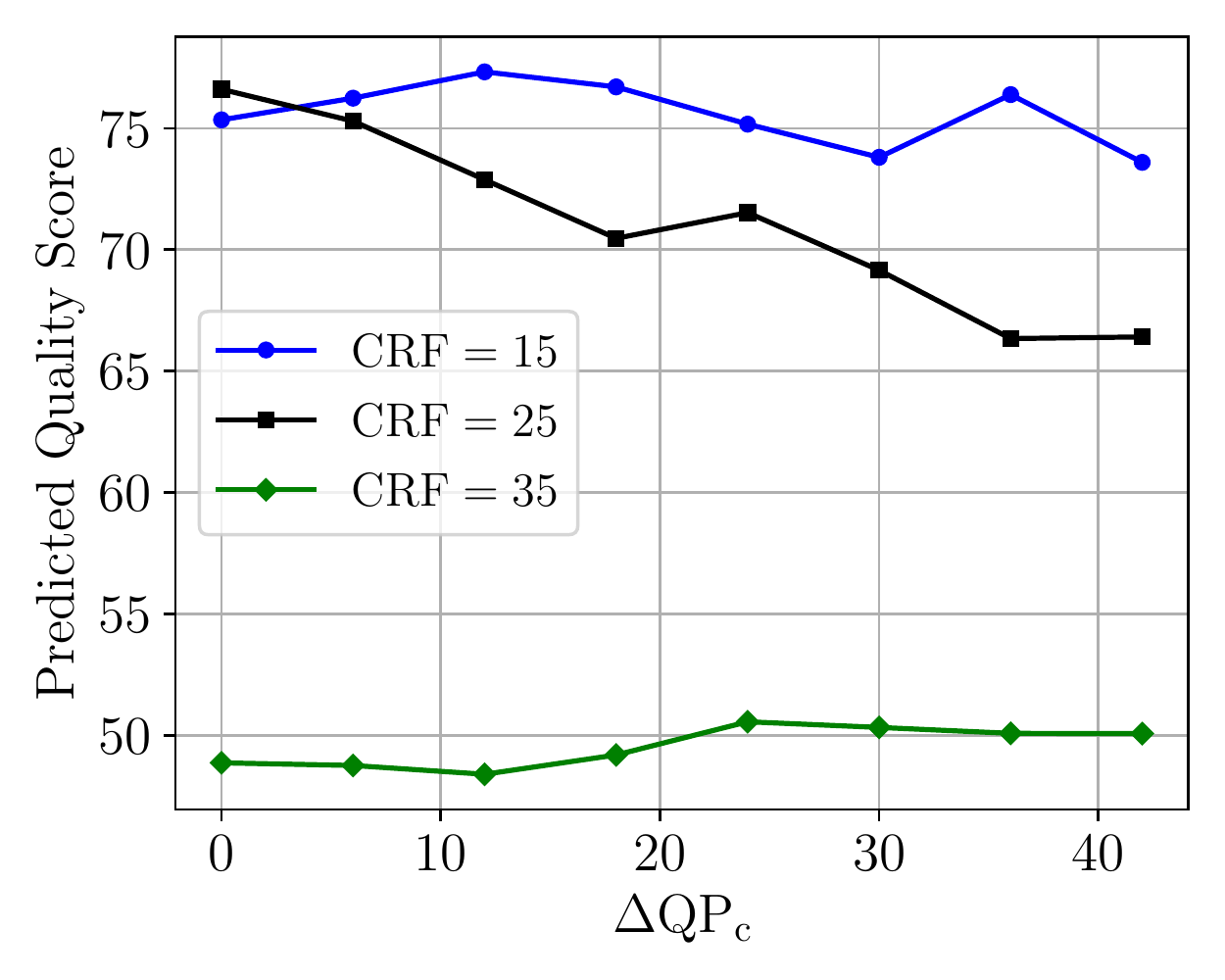} &
    \includegraphics[width=\imgwid]{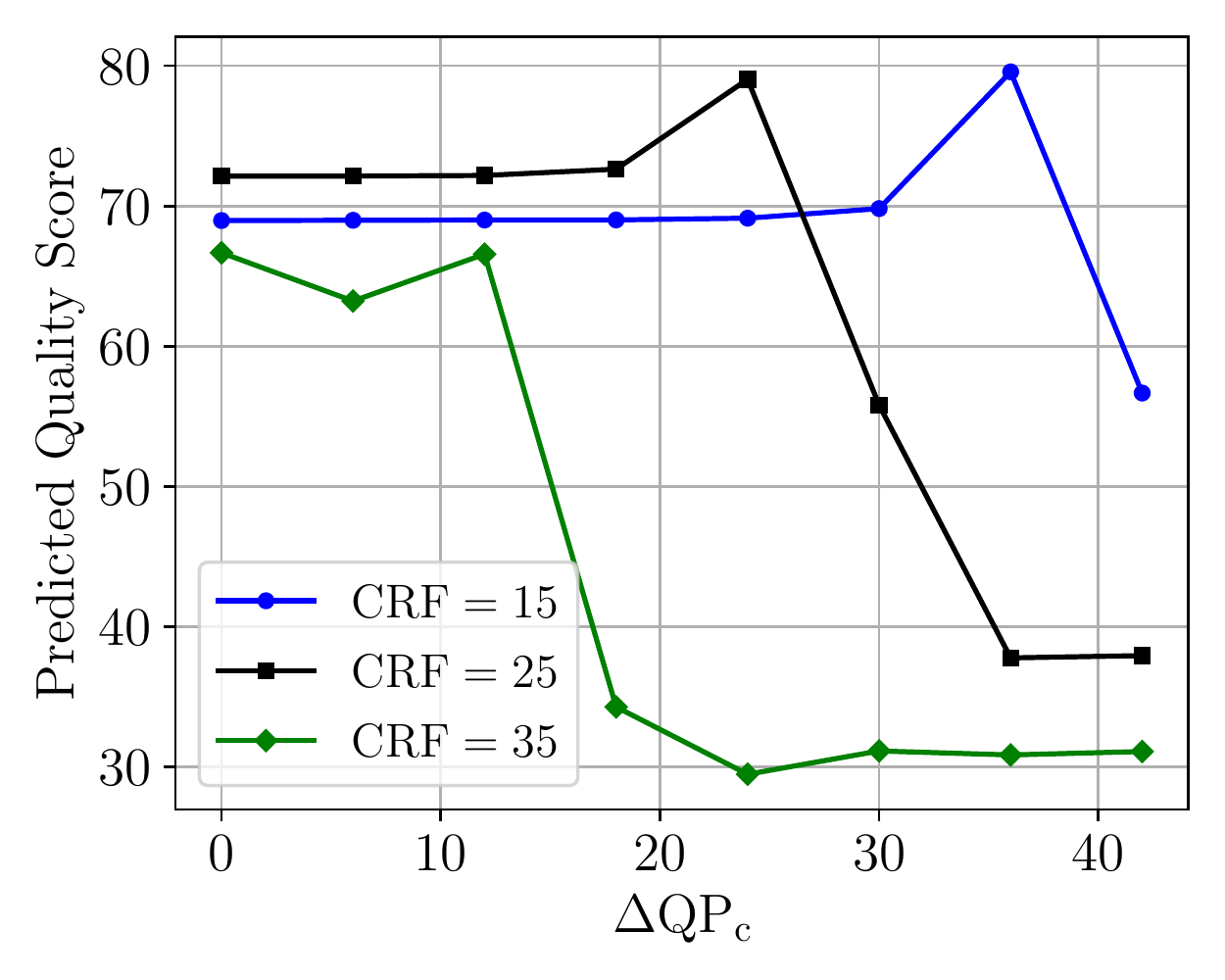} \\
    (c) Fail model (avg.) & (d) Fail model (\textit{blue\_sky\_1080p25})
  \end{tabular}
  \caption{Monotonicity analysis for variants of trained VMAF models. (a) and (b) are the predictions from VMAF 0.6.1 and VMAF$_\text{c}$ (averaged over 15 sequences), respectively. (c) and (d) show failed cases of a model trained with $(C,\gamma)=(2^3,2^2)$ using the same features as VMAF$_\text{c}$, but without quantization.}
  \label{fig:mono_plots}
\end{figure}

\subsection{Monotonicity Analysis}
\label{sec:mono}
Lastly, we study the monotonicity property of the trained models. Ideally, a VQA model should satisfy the following property: when the \textit{chroma\_qp\_offset} parameter is increased, while fixing the other parameters, the predicted quality score $M$ should be monotonically non-increasing. That is, given two \textit{chroma\_qp\_offset} values $\Delta \mathrm{QP_{c,1}}\leq \Delta \mathrm{QP_{c,2}}$, then we desire that $M_1 \ge M_2$. Similarly, if \textit{chroma\_qp\_offset} is fixed, the predicted quality score should decrease or maintain at the same level, as the CRF is increased. This would allow the model to be used for constructing bitrate ladders and for calculating BD-rate.

We collected 15 test video contents of 1080p resolution and YUV420 format from Xiph Video Test Media\footnote[5]{https://media.xiph.org/video/derf/} to conduct the experiment. The source videos were encoded at 3 different CRF values: 15, 25, and 35. At each CRF level, we further assigned 8 equally separated $\Delta\mathrm{QP_c}$ steps, ranging from 0 to 42. As shown in Fig. \ref{fig:mono_plots}, VMAF 0.6.1 delivered a flat result with respect to $\Delta \mathrm{QP_{c}}$, as should be expected, since its features only ingest luma information. By contrast, the perfectly monotonic plot given by VMAF$_\text{c}$ indicates that the additional chromatic features were properly integrated. It may be observed in Fig. \ref{fig:mono_plots}(b) that the plot for $\mathrm{CRF}=35$ saturates fast. This is because the quantization parameter in (\ref{eq:finalCQP}) already reaches its maximum value of $51$ at $\Delta \mathrm{QP_{c}}=24$. We also demonstrated the merit of this analysis by a fail case in Figs. \ref{fig:mono_plots}(c), \ref{fig:mono_plots}(d): this model achieves $0.730$ of SROCC on the NFLX$_\text{c}$ dataset, which is a substantial improvement as compared with the $0.612$ of SROCC of VMAF 0.6.1. Unfortunately, the monotonicity property does not hold individually or on average, which suggests the possibility of overfitting.

\section{Conclusion and Future Work}
We have constructed a subjective video quality study and database to support the design of algorithms that can better predict the quality of chromatically distorted videos. This new resource contains HEVC-compressed video contents, spanning wide ranges of quality levels applied differently on a per-channel basis. We also improved an existing high-performing, learning based VQA model (VMAF) by integrating two simple features extracted from chroma channels, and compared the performance of the new chroma-sensitized model against several leading objective VQA metrics. The new VMAF$_\text{c}$ model was found to be the top performer on the new chroma distortion dataset NFLX$_\text{c}$.

We hope that this work encourages increased awareness of chromatic distortions in the design of quality models, databases, and future compression protocols. The results from our human study indicate that there is room for improving perceptual coding efficiency in modern video codecs, by increasing the compression factor on chroma channels. With the improved video quality model, it is possible to jointly optimize luma and chroma compression in video encoders. For example, the chroma components could be further subsampled or quantized without suffering perceptual fidelity.

Next, we plan to seek new chromatic features that can be used to both effectively capture chroma distortions, as well as preserve performance on luminance distortions. Investigating more sophisticated machine learning models is also of interest. Although compression engines create the most prevalent and common artifacts in streaming video scenarios, there exist other important sources of chromatic distortions, such as subsampling and chroma noise. Creating databases and VQA algorithms that address different kinds of realistic chromatic distortions would be quite valuable. New distortion types emerging from deep video processing problems \cite{liu2019cyclicgen,Ko2020} are also worthy of study. Looking further ahead, developing protocols to optimize video encoders by exploiting improved models of chromatic distortion perception is an intriguing topic of potentially high practical impact.




\section*{Acknowledgment}
The authors would like to acknowledge Andrey Norkin and Mariana Afonso for fruitful discussions on subjective study design. The authors also thank the Video Algorithms team at Netflix for supporting this work.

\ifCLASSOPTIONcaptionsoff
  \newpage
\fi


\bibliographystyle{IEEEtran}
\bibliography{bare_jrnl}{}








\end{document}